\documentclass[11pt, prd, reprint,showkeys, amsmath,amssymb, aps, nofootinbib, floatfix]{revtex4-2}
\DeclareUnicodeCharacter{2212}{\textendash}
\usepackage{graphicx,subfigure}
\graphicspath{{figures/}}
\usepackage[export]{adjustbox}
\usepackage{dcolumn}
\usepackage[outdir=./epsTopdf/]{epstopdf}
\usepackage{bm}
\usepackage{xcolor}
\usepackage{slashed,cancel}
\usepackage{multirow}
\usepackage{float}
\usepackage{capt-of}
\usepackage{comment}
\usepackage{mathrsfs}
\usepackage{tikz-feynman}
\usepackage{enumitem}
\usepackage[colorlinks=true,linkcolor=blue,urlcolor=blue,citecolor=blue]{hyperref}
\newcommand{\prlsection}[1]{\noindent\textbf{#1}}
\allowdisplaybreaks

\begin{document}

\title{High energy probes of Higgs self-coupling via $W$ boson fusion at future lepton colliders}

\author{Amir Subba}
\email{amirsubba@ustc.edu.cn}
\affiliation{Wilczek Quantum Center, Shanghai Institute for Advanced Studies, Shanghai 201315, China}
\affiliation{University of Science and Technology of China, Hefei 230026, China}
\author{Hrishikesh Deka}
\email{hrishikesh.deka@iitg.ac.in}
\affiliation{Department of Physics, Indian Institute of Technology Guwahati, Guwahati, Assam, 781039, India}
\author{Subhaditya Bhattacharya}
\email{subhab@iitg.ac.in}
\affiliation{Department of Physics, Indian Institute of Technology Guwahati, Guwahati, Assam, 781039, India}
\author{Abhik Sarkar}
\email{asarkar@lpsc.in2p3.fr}
\affiliation{Department of Physics, Indian Institute of Technology Guwahati, Guwahati, Assam, 781039, India}
\affiliation{Laboratoire de Physique Subatomique et de Cosmologie, Université Grenoble-Alpes, CNRS/IN2P3, 53 Avenue des Martyrs, 38026 Grenoble, France}

\begin{abstract}
We investigate the sensitivity to the Higgs self-coupling through $W$ boson
fusion di-Higgs production at CLIC with a center-of-mass energy of
$\sqrt{s}=3$~TeV. We study the interplay between the Higgs self-coupling
modifier ($\kappa_{\lambda}$) and Higgs-gauge boson coupling modifiers ($\kappa_{V}$
and $\kappa_{2V}$) within the $\kappa$ framework. To enhance the separation between signal and background, we develop
a graph neural network (GNN) based classifier that achieves a signal significance of
$\mathscr{Z}\approx 20~\sigma$ at $5~\mathrm{ab}^{-1}$, substantially exceeding
projected HL-LHC sensitivity. Our results demonstrate that high-energy lepton
colliders, combined with graph-based machine learning, provide excellent
sensitivity to  the Higgs self-coupling and offer a powerful probe of new physics in the electroweak sector, potentially disentangling underlying SMEFT-like and HEFT-like structures. 
\end{abstract}

\maketitle

\section{Introduction}
\label{sec:intro}
The discovery of Higgs boson at the LHC \cite{ATLAS:2012yve, CMS:2012qbp} provides a long sought-after probe of the spontaneous symmetry breaking of the electroweak sector described by the $SU(2)_L\times U(1)_Y$ gauge theory of the Standard Model (SM). The mechanism of electroweak symmetry breaking (EWSB)~\cite{Higgs:1964pj,Englert:1964et,Guralnik:1964eu,Higgs:1964ia,Higgs:1966ev,Kibble:1967sv} is encoded in the Higgs potential $V(H)=\mu^{2}(H^{\dagger}H)+\lambda(H^{\dagger}H)^{2}$,\footnote{In SM, $H$ represents a Higgs doublet in unitary gauge which takes the form $\frac{1}{\sqrt {2}} \begin{pmatrix}
    0\\v+h
\end{pmatrix}$} where the self-coupling $\lambda$ governs the shape of the potential and directly determines the trilinear and quartic Higgs self-interactions. Its precise measurement is therefore central to validating the SM scalar sector and probing the nature of EWSB. A direct determination of $\lambda$ requires di-Higgs production, whose cross section is small and the signal phase space must be disentangled from large SM backgrounds, a challenge that has driven extensive phenomenological and experimental activity at the LHC~\cite{ATLAS:2024ish,CMS:2025ngq} and motivates projections for the High-Luminosity LHC (HL-LHC)~\cite{ATLAS:2022faz,ATLAS:2025cwr,Kim:2018uty,Brivio:2025sib,Adhikary:2017jtu}, as well as future lepton colliders~\cite{List:2024ukv,Torndal:2023mmr,Buonincontri:2022ylv,Cheung:2026vdk,Roloff:2019crr}.

The current work investigates the sensitivity of the future 3~TeV $e^+e^-$ run of the Compact Linear Collider (CLIC) to the Higgs self-coupling. Lepton colliders offer a clean experimental environment with negligible QCD backgrounds, making them ideal for precision studies of the electroweak sector. At multi-TeV center-of-mass (CM) energies, $W$ boson
fusion (WBF) becomes the dominant di-Higgs production mechanism. We focus on the channel $e^+e^-\to hh\slashed{E}$, where $\slashed{E}$ stands for missing energy, as it provides the greatest sensitivity to the Higgs self-coupling at CLIC, with results directly applicable to future $\mu^+\mu^-$ colliders operating at comparable CM energies. 

The WBF process receives contributions from three leading diagrams involving the trilinear Higgs self-coupling and the Higgs-gauge boson interactions, whose interference significantly influences the sensitivity to $\kappa_\lambda$, the modifier to the Higgs self-coupling. The nature of the underlying new physics (NP) that modifies these interactions determines whether deviations in $\kappa_V$ and $\kappa_{2V}$, the Higgs-gauge boson coupling modifiers,
are correlated, as in SMEFT~\cite{Buchmuller:1985jz,Grzadkowski:2010es}, or independent, as in HEFT~\cite{Feruglio:1992wf,Alonso:2012px,Burgess:1999ha,Contino:2010mh,Buchalla:2012qq,Buchalla:2015wfa}. Di-Higgs production thus provides both a direct probe of the Higgs self-interaction and an indirect window into the structure of the Higgs sector beyond the SM.

A central challenge in this analysis is discriminating the small di-Higgs signal from overwhelming SM backgrounds. We address this with a two-stage strategy: optimized selection cuts first suppress reducible backgrounds while preserving signal efficiency, followed by a graph neural network (GNN) serving as the primary signal-background discriminant. GNNs are
particularly well-suited to collider events, which naturally admit a graph representation with jets as nodes and angular proximity as edges, allowing the network to exploit both local kinematic correlations and global event topology simultaneously.

Higgs self coupling at CLIC has already been estimated~\cite{Roloff:2019crr} showing a significant betterment than the HL-LHC prediction. In this work, however, we point out several theoretical and phenomenological implications like the effect of the interference between Higgs-gauge and Higgs self coupling diagrams, the advantage of Graph-based Neural Network on optimizing the significance, the possibility of disentangling HEFT and SMEFT frameworks,  the statistical estimates of the coupling modifiers at CLIC environment, etc.

\section{Trilinear Higgs Coupling}
\label{sec:3Higgs}
Following the convention adopted in LHC analyses, we parameterize possible
NP effects in the Higgs sector using the $\kappa$ framework~\cite{ATLAS:2022vkf,CMS:2022dwd}. Deviations in the Higgs self-interactions are described by the trilinear and quartic coupling modifiers,
\begin{equation}
    \kappa_{\lambda} =
    \frac{\lambda^{hhh}_{\rm obs}}{\lambda^{hhh}_{\rm SM}},
    \qquad
    \kappa_{2\lambda} =
    \frac{\lambda^{hhhh}_{\rm obs}}{\lambda^{hhhh}_{\rm SM}},
\end{equation}
Deviations in the Higgs couplings to electroweak gauge bosons are
parameterized by
\begin{equation}
    \kappa_V =
    \frac{g^{hVV}_{\rm obs}}{g^{hVV}_{\rm SM}},
    \qquad
    \kappa_{2V} =
    \frac{g^{hhVV}_{\rm obs}}{g^{hhVV}_{\rm SM}}.
\end{equation}
Here $\lambda^{hhh}$ ($\lambda^{hhhh}$) and $g^{hVV}$ ($g^{hhVV}$) denote
the trilinear (quartic) Higgs self-couplings and the single (double) Higgs
couplings to electroweak gauge bosons, respectively. Since the quartic
modifier $\kappa_{2\lambda}$ contributes only through highly suppressed
processes, it is neglected in the processes that we consider here. We focus instead on the leading dependence
of di-Higgs production on $\kappa_{\lambda}$ and $\kappa_{2V}$, with
$\kappa_V$ treated under two representative scenarios:
\begin{itemize}
    \itemsep-0em
    \item \textit{Case 1}: $\kappa_V=1$, with $\kappa_{\lambda}$ and
    $\kappa_{2V}$ varied independently. This is motivated by non-linearly
    realized EWSB frameworks such as HEFT, where $hVV$ and $hhVV$
    interactions are independent.
    \item \textit{Case 2}: $\kappa_V=\kappa_{2V}$, with $\kappa_{\lambda}$
    and $\kappa_{2V}$ as free parameters. This relation is characteristic
    of SMEFT-like frameworks where both interactions arise from EWSB and
    are correlated. (see App.~\ref{appA} for details).
\end{itemize}

At hadron colliders, the dominant di-Higgs production modes include gluon-gluon fusion, vector boson fusion, and associated production with a top-quark pair or an electroweak gauge boson. At lepton colliders, the most important channels are $Z$-associated production and VBF. The
strongest current constraints on $\kappa_\lambda (\kappa_{\rm 2V})$ are $[-1.2, 7.2]$ ($[0.6, 1.5]$) from ATLAS~\cite{ATLAS:2024ish} and $[-1.35, 6.37]$ ($[0.64, 1.40]$) from CMS~\cite{CMS:2025ngq} at 95\% CL. At the HL-LHC at $3~\mathrm{ab}^{-1}$, sensitivity of $[0.3,1.9]$ ($[0.8,1.2]$) is expected without systematics at 95\% C.L~\cite{ATLAS:2022faz,ATLAS:2025cwr}. $\kappa_{V}$ has been extensively constrained across multiple studies at the LHC, and the combined average is $1.023\pm0.026$~\cite{ParticleDataGroup:2026aaa}.

\section{Di-Higgs Production at CLIC}
\label{sec:Di-Higgs}
\begin{figure}[htb!]
    \centering
    \includegraphics[width=\linewidth]{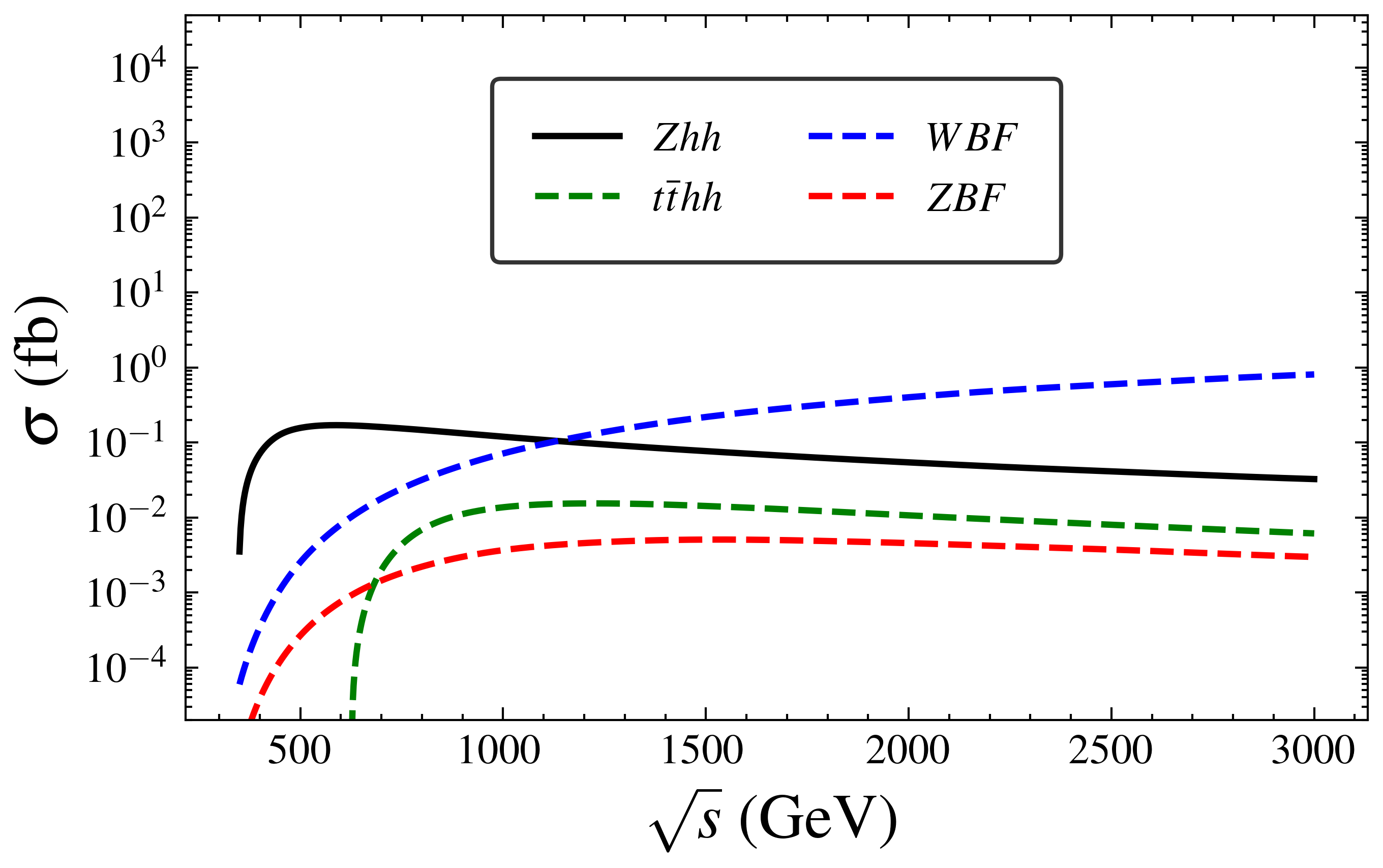}
    \caption{Di-Higgs production cross sections as a function of CM energy
    $\sqrt{s}$ at $e^+e^-$ colliders.}
    \label{fig:dihxs}
\end{figure}

Di-Higgs production at lepton colliders proceeds through several channels,
namely $Zhh$, WBF, $Z$-boson fusion (ZBF), and $t\bar{t}hh$. The cross
sections as a function of CM energy are shown in Fig.~\ref{fig:dihxs}. At low CM energies the $Zhh$ process dominates, but
being an $s$-channel process its cross section falls rapidly with
increasing CM energy. The ZBF and $t\bar{t}hh$ channels remain subdominant
throughout. For CM energies $\gtrsim 1$~TeV, WBF becomes the dominant
di-Higgs production mechanism, making it the primary focus of our analysis
at the 3~TeV CLIC. The contribution from $Z(\nu\bar{\nu})hh$ to the same
final state is more than two orders of magnitude smaller and is neglected. 

\begin{figure*}[htb!]    
    {\Large
    \subfigure[]{\begin{tikzpicture}[baseline={(current bounding box.center)},style={scale=0.5, transform shape}]
	\begin{feynman}
		\vertex(a);
		\vertex[above left =0.75cm and 2.25cm of a] (a1){$\boldsymbol{e^{-}}$};
        \vertex[below = 1.5cm of a] (c);
        \vertex[right = 1.25cm of c] (d);
        \vertex[below = 1.5cm of c] (b);
		\vertex[below left =0.75cm and 2.25cm of b] (b1){$\boldsymbol{e^{+}}$};
        \vertex[above right =0.75cm and 2.25cm of a] (a2){$\boldsymbol{\nu_{e}}$};
        \vertex[below right =0.75cm and 2.25cm of b] (b2){$\boldsymbol{\overline{\nu}_{e}}$};
        \vertex[above right =0.75cm and 1cm of d] (c1){$\boldsymbol{h}$};
        \vertex[below right =0.75cm and 1cm of d] (c2){$\boldsymbol{h}$};
		\diagram*{
			(a1) -- [thick, fermion, arrow size=1.25pt] (a),
                (a) -- [thick, boson, edge label' = $\boldsymbol{W}$, arrow size=1.25pt] (c),
                (b) -- [thick, boson, edge label = $\boldsymbol{W}$, arrow size=1.25pt] (c),
			(b) -- [thick, fermion, arrow size=1.25pt] (b1),
                (a) -- [thick, fermion, arrow size=1.25pt] (a2),
                (b2) -- [thick, fermion, arrow size=1.25pt] (b),
                (d) -- [thick, scalar, edge label = $\boldsymbol{h}$, arrow size=1.25pt] (c),
                (d) -- [thick, scalar, arrow size=1.25pt] (c1),
                (d) -- [thick, scalar, arrow size=1.25pt] (c2)
		};
	\end{feynman}
    \end{tikzpicture}
    \label{fig:2h2vA}}
    \subfigure[]{\begin{tikzpicture}[baseline={(current bounding box.center)},style={scale=0.5, transform shape}]
	\begin{feynman}
		\vertex(a);
		\vertex[above left =0.75cm and 2.25cm of a] (a1){$\boldsymbol{e^{-}}$};
        \vertex[below = 1.5cm of a] (c);
        \vertex[below = 1.5cm of c] (b);
		\vertex[below left =0.75cm and 2.25cm of b] (b1){$\boldsymbol{e^{+}}$};
        \vertex[above right =0.75cm and 2.25cm of a] (a2){$\boldsymbol{\nu_{e}}$};
        \vertex[below right =0.75cm and 2.25cm of b] (b2){$\boldsymbol{\overline{\nu}_{e}}$};
        \vertex[above right =0.75cm and 2.25cm of c] (c1){$\boldsymbol{h}$};
        \vertex[below right =0.75cm and 2.25cm of c] (c2){$\boldsymbol{h}$};
		\diagram*{
			(a1) -- [thick, fermion, arrow size=1.25pt] (a),
                (a) -- [thick, boson, edge label' = $\boldsymbol{W}$, arrow size=1.25pt] (c),
                (b) -- [thick, boson, edge label = $\boldsymbol{W}$, arrow size=1.25pt] (c),
			(b) -- [thick, fermion, arrow size=1.25pt] (b1),
                (a) -- [thick, fermion, arrow size=1.25pt] (a2),
                (b2) -- [thick, fermion, arrow size=1.25pt] (b),
                (c) -- [thick, scalar, arrow size=1.25pt] (c1),
                (c) -- [thick, scalar, arrow size=1.25pt] (c2)
		};
	\end{feynman}
	\end{tikzpicture}
        \label{fig:2h2vB}}
    \subfigure[]{\begin{tikzpicture}[baseline={(current bounding box.center)},style={scale=0.5, transform shape}]
	\begin{feynman}
		\vertex(a);
		\vertex[above left =0.75cm and 2.25cm of a] (a1){$\boldsymbol{e^{-}}$};
        \vertex[below = 1.0cm of a] (c);
        \vertex[below = 1.0cm of c] (d);
        \vertex[below = 1.0cm of d] (b);
		\vertex[below left =0.75cm and 2.25cm of b] (b1){$\boldsymbol{e^{+}}$};
        \vertex[above right =0.75cm and 2.25cm of a] (a2){$\boldsymbol{\nu_{e}}$};
        \vertex[below right =0.75cm and 2.25cm of b] (b2){$\boldsymbol{\overline{\nu}_{e}}$};
        \vertex[above right =0.25cm and 2.25cm of c] (c1){$\boldsymbol{h}$};
        \vertex[below right =0.25cm and 2.25cm of d] (c2){$\boldsymbol{h}$};
		\diagram*{
			(a1) -- [thick, fermion, arrow size=1.25pt] (a),
                (a) -- [thick, boson, edge label' = $\boldsymbol{W}$, arrow size=1.25pt] (c),
                (c) -- [thick, boson, edge label' = $\boldsymbol{W}$, arrow size=1.25pt] (d),
                (b) -- [thick, boson, edge label = $\boldsymbol{W}$, arrow size=1.25pt] (d),
			(b) -- [thick, fermion, arrow size=1.25pt] (b1),
                (a) -- [thick, fermion, arrow size=1.25pt] (a2),
                (b2) -- [thick, fermion, arrow size=1.25pt] (b),
                (c) -- [thick, scalar, arrow size=1.25pt] (c1),
                (d) -- [thick, scalar, arrow size=1.25pt] (c2)
		};
	\end{feynman}
	\end{tikzpicture}
        \label{fig:2h2vC}}
    \subfigure[]{\begin{tikzpicture}[baseline={(current bounding box.center)},style={scale=0.5, transform shape}]
	\begin{feynman}
		\vertex(a);
		\vertex[above left =0.75cm and 2.25cm of a] (a1){$\boldsymbol{e^{-}}$};
        \vertex[below = 1.0cm of a] (c);
        \vertex[below = 1.0cm of c] (d);
        \vertex[below = 1.0cm of d] (b);
		\vertex[below left =0.75cm and 2.25cm of b] (b1){$\boldsymbol{e^{+}}$};
        \vertex[above right =0.75cm and 2.25cm of a] (a2){$\boldsymbol{\nu_{e}}$};
        \vertex[below right =0.75cm and 2.25cm of b] (b2){$\boldsymbol{\overline{\nu}_{e}}$};
        \vertex[above right =0.25cm and 2.25cm of c] (c1){$\boldsymbol{h}$};
        \vertex[below right =0.25cm and 2.25cm of d] (c2){$\boldsymbol{h}$};
		\diagram*{
			(a1) -- [thick, fermion, arrow size=1.25pt] (a),
                (a) -- [thick, boson, edge label' = $\boldsymbol{W}$, arrow size=1.25pt] (c),
                (c) -- [thick, boson, edge label' = $\boldsymbol{W}$, arrow size=1.25pt] (d),
                (b) -- [thick, boson, edge label = $\boldsymbol{W}$, arrow size=1.25pt] (d),
			(b) -- [thick, fermion, arrow size=1.25pt] (b1),
                (a) -- [thick, fermion, arrow size=1.25pt] (a2),
                (b2) -- [thick, fermion, arrow size=1.25pt] (b),
                (c) -- [thick, scalar, arrow size=1.25pt] (c2),
                (d) -- [thick, scalar, arrow size=1.25pt] (c1)
		};
	\end{feynman}
	\end{tikzpicture}
        \label{fig:2h2vD}}}
    \caption{ \label{fig:dihiggs} Feynman diagrams
    contributing to di-Higgs production via WBF.}
    
\end{figure*}
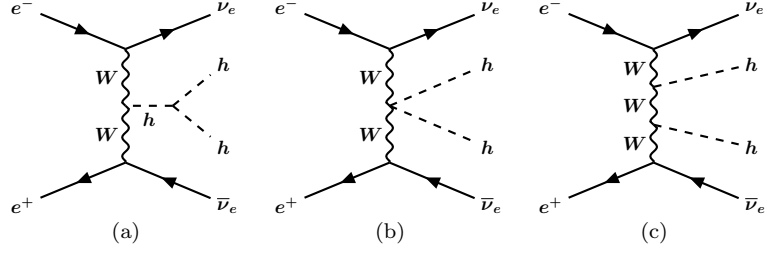
The three leading Feynman diagrams to di-Higgs plus $\slashed{E}$ production are shown in the Fig.~\ref{fig:dihiggs}. Diagram~\ref{fig:2h2vA} contains the trilinear Higgs self-coupling with amplitude proportional to $\kappa_V\kappa_\lambda$. Diagrams~\ref{fig:2h2vB} and~\ref{fig:2h2vC} involve only Higgs-gauge boson interactions, with amplitudes proportional to $\kappa_{2V}$ and $\kappa_V^2$,
respectively. While the amplitudes of Figs.~\ref{fig:2h2vA} and~\ref{fig:2h2vB} interfere constructively, both interfere destructively with Fig.~\ref{fig:2h2vC}. The explicit cross section dependence on the coupling modifiers is provided in App.~\ref{appB}. Since the interference between the $\kappa_{2V}$ and $\kappa_V^2$ amplitudes is particularly strong, the projected sensitivity to $\kappa_{2V}$ differs significantly between the two benchmark scenarios. \\

\prlsection{Event Classification}
We focus on $4b+\slashed{E}$ signal here, stemming from the process: $e^+e^- \rightarrow hh\nu \bar \nu,~h\rightarrow b\bar b,~h\rightarrow b\bar b$. The dominant backgrounds are single-Higgs associated production ($hb\bar{b}\slashed{E}$),  multi-$b$ production ($4b\slashed{E}$), and top-quark processes ($t\bar{t}X$). Signal and background events are simulated at leading order with \texttt{MG5\_aMC$@$NLO}~\cite{Alwall:2014hca} at $\sqrt{s}=3$~TeV, hadronized with \texttt{Pythia8}~\cite{Bierlich:2022pfr}, and passed through \texttt{Delphes3}~\cite{deFavereau:2013fsa} for detector simulation. Isolated leptons with $p_T>10$~GeV and anti-$k_T$ ($R=0.4$)~\cite{Cacciari:2008gp} jets with $p_T>20$~GeV are retained. Events must satisfy $N_j\ge 4$, $N_b\ge 2$, and $\slashed{E}_T\ge 20$~GeV, where $N_j$ and $N_b$ denote the number of reconstructed jets and $b$-tagged jets, respectively, $\slashed{E}_T$ is the missing transverse energy. Events with isolated leptons ($N_\ell\ne 0$) are vetoed. Among
the four leading jets, all three disjoint pair combinations are formed and an event is retained only if at least one pair yields a di-jet invariant mass within $m_h\pm 25$~GeV $(m_h = 125~\mathrm{GeV})$; this Higgs mass-window requirement renders the $4b\slashed{E}$ contribution negligible.

The surviving events are classified using a message-passing GNN with ${\approx}\,2.6\times10^5$ trainable parameters, detailed in App.~\ref{appC}. The classifier achieves near-ideal separation of $t\bar{t}X$ from the signal (AUC $=0.999$) and an overall micro-averaged AUC of $0.95$. For the SM benchmark $\kappa_\lambda=\kappa_{2V}=\kappa_V=1$, the post-classification signal and background efficiencies are $\epsilon_S\simeq 2.0\times10^{-2}$ and
$\epsilon_B\simeq 6\times10^{-4}$, respectively, at $P(hh\slashed{E})\ge 0.5$, with $P(X)$ being the probability for events to be tagged as signal. Using the profile-likelihood significance~\cite{ParticleDataGroup:2026aaa}
\begin{equation}
\label{eq:z}
    \mathscr{Z}=
    \sqrt{2\left[(S+B)\ln\!\left(1+\frac{S}{B}\right)-S\right]},
\end{equation}
we obtain $\mathscr{Z}\approx 8$, $15$, and $20$ at integrated luminosities of $1$, $3$, and $5~\mathrm{ab}^{-1}$, respectively, establishing discovery-level sensitivity well within the nominal CLIC luminosity program, see App.~\ref{appD} for details.

\section{Sensitivity of Couplings}
\label{sec:sens}
\begin{figure*}[htb!]
\centering
\includegraphics[width=0.44\linewidth]{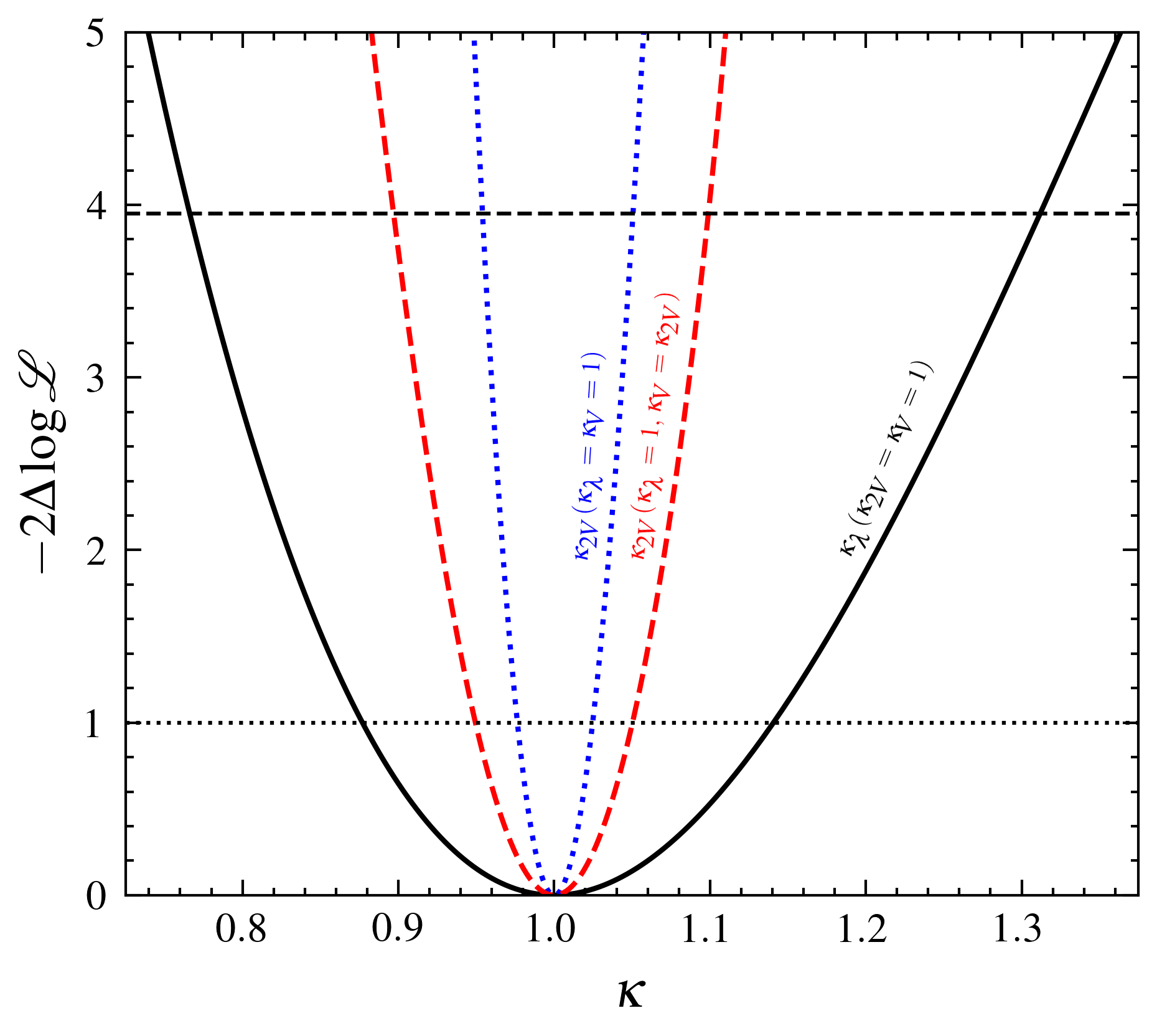}
\includegraphics[width=0.45\linewidth]{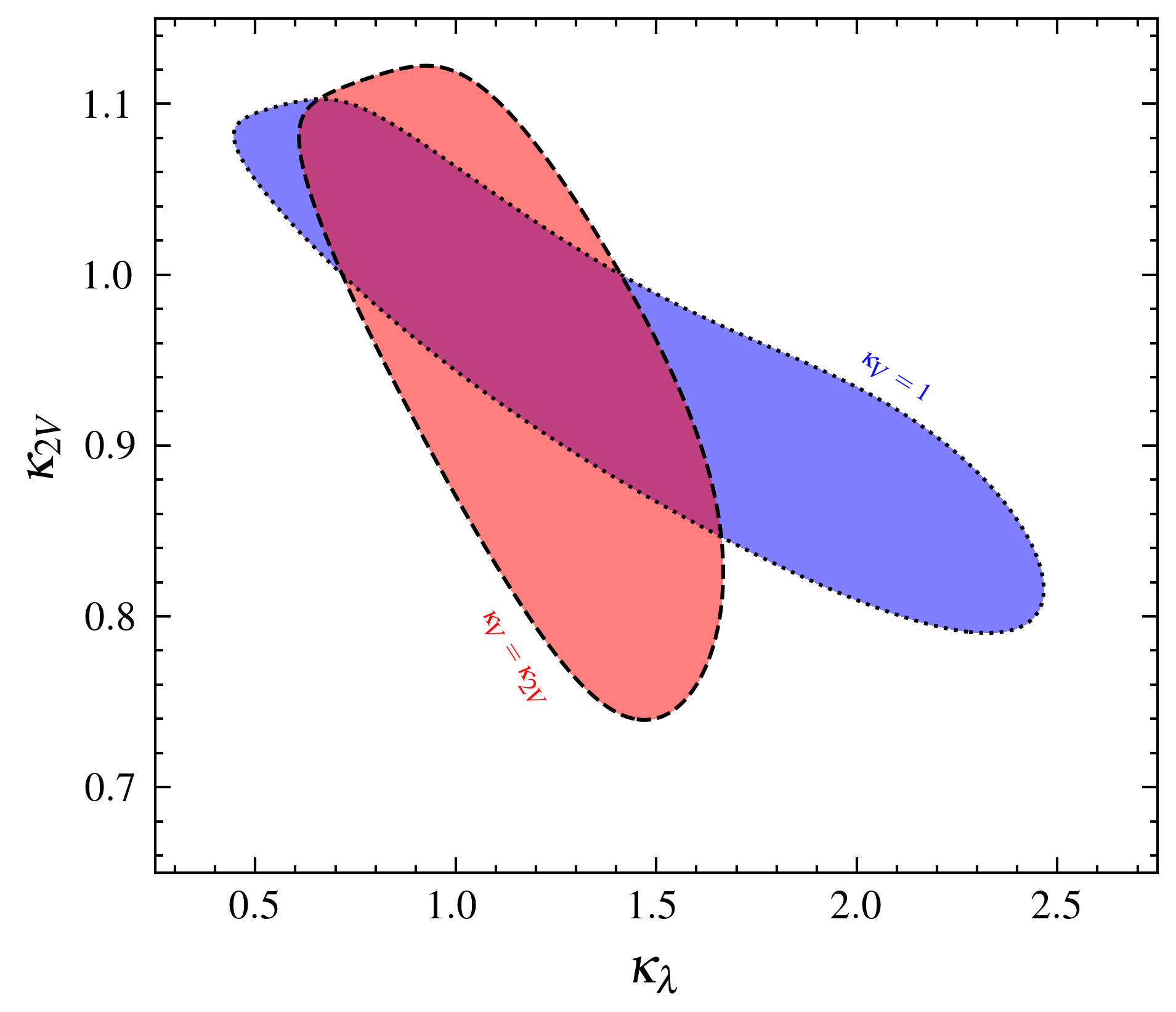}
\caption{1D likelihood scans (\textit{left}) and $95\%$ confidence level. 2D
sensitivity contours (\textit{right}) at CLIC $\sqrt{s}=3$~TeV with
$\mathfrak{L}_{\rm int}=5~\mathrm{ab}^{-1}$.}
\label{fig:sens}
\end{figure*}
We consider five equally spaced bins for each of the observables $m_{hh}\in[0.2,1.9]~\mathrm{TeV}$, $\Delta\eta_{hh}\in[0,7.5]$, and $\Delta\phi_{hh}\in[-\pi,\pi]$. The reconstruction of the two Higgs bosons uses the four leading jets in each event. Since there are three ways to partition four jets into two di-jet systems, namely $(j_1,j_2;j_3,j_4)$, $(j_1,j_3;j_2,j_4)$, and $(j_1,j_4;j_2,j_3)$, all three combinations are
considered and the optimal pairing is selected by minimizing the $\chi^2$ statistic
\begin{equation}
\chi^2 =
\frac{(\Delta m_{h,1})^2+(\Delta m_{h,2})^2}{\sigma_h^2} +
\frac{(\Delta m_{1,2})^2}{\sigma_\Delta^2},
\end{equation}
where $\Delta m_{h,1(2)}=m^{jj}_{1(2)}-m_h$, $\Delta m_{1,2}=m^{jj}_{1}-m^{jj}_{2}$, $m_h=125$~GeV, and $\sigma_h$, $\sigma_\Delta$ characterize the di-jet mass and mass-difference
resolutions, respectively. The four-momenta of the reconstructed Higgs bosons are then used to construct the kinematic observables above.

To quantify the sensitivity to the anomalous couplings, we perform a binned likelihood analysis~\cite{ParticleDataGroup:2026aaa}. The expected number of events in bin $i$ is
\begin{equation}
\mu_i(\kappa)=s_i(\kappa)+b_i(\kappa),
\end{equation}
with the SM Asimov dataset $n_i=s_i(\kappa=1)+b_i(\kappa=1)$. The likelihood function
\begin{equation}
\mathscr{L}(\kappa)= \prod_i \frac{\mu_i(\kappa)^{n_i}e^{-\mu_i(\kappa)}}{n_i!}
\end{equation}
yields the test statistic
\begin{equation}
\mathcal{Q}(\kappa)
=2\sum_i
\left[
\mu_i(\kappa)-n_i
+n_i\ln\!\left(\frac{n_i}{\mu_i(\kappa)}\right)
\right].
\end{equation}
The resulting 1D likelihood scans at $\mathfrak{L}_{\rm int}=5~\mathrm{ab}^{-1}$ are shown in Fig.~\ref{fig:sens} (\textit{left} panel); results at lower luminosities are given in App.~\ref{appE}. The likelihood profile for $\kappa_\lambda$ exhibits a well-defined minimum at the SM expectation, $\kappa_\lambda=1$, with $68\%$ and $95\%$ confidence intervals,
\begin{equation}
\kappa_{\lambda}=
\begin{cases}
[0.88,\,1.14] &(68\%~\mathrm{C.L.})\\
[0.76,\,1.31] &(95\%~\mathrm{C.L.})
\end{cases}.
\end{equation}
The sensitivity to $\kappa_{2V}$ is considerably stronger, reflecting its pronounced impact on both the total production rate and the kinematic properties of the di-Higgs system. For \textit{Case 1} ($\kappa_V=1$),
\begin{equation}
\kappa_{2V}=
\begin{cases}
[0.98,\,1.02] &(68\%~\mathrm{C.L.})\\
[0.95,\,1.05] &(95\%~\mathrm{C.L.})
\end{cases},
\end{equation}
whereas imposing \textit{Case 2} ($\kappa_V=\kappa_{2V}$) yields the weaker constraints
\begin{equation}
\kappa_{2V}=
\begin{cases}
[0.95,\,1.05] &(68\%~\mathrm{C.L.})\\
[0.90,\,1.10] &(95\%~\mathrm{C.L.})
\end{cases}.
\end{equation}
The broadening of the allowed interval in \textit{Case 2} directly reflects the correlation between $\kappa_V$ and $\kappa_{2V}$, highlighting the importance of a simultaneous fit of these couplings in di-Higgs production. The sensitivities presented here include only statistical uncertainties while the impact of systematic uncertainties is assessed in App.~\ref{appE}.

The 2D correlated likelihood contours in the $\kappa_{\lambda}$-$\kappa_{2V}$ plane are shown in the right panel of Fig.~\ref{fig:sens}. In \textit{Case 1}, the contours exhibit a pronounced correlation between $\kappa_{\lambda}$ and $\kappa_{2V}$. Imposing $\kappa_V=\kappa_{2V}$ in \textit{Case 2} modifies the interference pattern, relaxing the constraints and changing the shape of the allowed region. The distinct sensitivities in the two scenarios demonstrate the importance of Higgs-gauge boson coupling assumptions and highlight the potential of future lepton colliders to discriminate between different realizations of the Higgs sector, namely SMEFT and HEFT.

To contextualize our results, we quantitatively compare our projections with the CLIC baseline study~\cite{Roloff:2019crr} (CLICdp henceforth). While our analysis projects $\kappa_\lambda \in [0.88, 1.14]$ at $68\%$ C.L., the CLICdp study primarily reports a relative uncertainty of $-8\%$ and $+11\%$ on the trilinear coupling evaluated at $68\%$ C.L. This difference in overall sensitivity stems from several methodological and phenomenological distinctions. Regarding channel content, our work isolates the $4b+\slashed{E}$ final state, whereas the CLICdp analysis uses boosted decision tree (BDT)-based classifier incorporating both the $b\bar{b}b\bar{b}$ and $b\bar{b}WW^*$ decay channels. Furthermore, our sensitivity relies exclusively on WBF at $\sqrt{s}=3$ TeV, while the optimal CLICdp constraints are achieved by combining the 3 TeV WBF differential analysis with $ZHH$ and WBF measurements from the 1.4 TeV energy stage. On the simulation front, our analysis utilizes a fast detector simulation coupled with a GNN classifier, whereas~\cite{Roloff:2019crr} employ a full \texttt{GEANT4}-based simulation~\cite{GEANT4:2002zbu} of the CLIC ILD detector model. Finally, regarding systematic uncertainties, our study explicitly profiles the impact of systematics up to $\sigma_{\rm sys} = 0.1$ (see App.~\ref{appE}), while the CLICdp analysis assumes systematic effects to be irrelevant compared to statistical uncertainties and therefore evaluates limits using statistics alone. To provide a fair comparison between the BDT and GNN-based classifiers, we present a comparison study in App.~\ref{app:bdt}.

\section{Conclusion}
\label{sec:conc}
We investigated the prospects for probing the Higgs self-coupling through WBF di-Higgs production at the 3~TeV CLIC with $5~\mathrm{ab}^{-1}$. Our results show substantial improvement over current LHC limits and projected HL-LHC sensitivity, underscoring the unique capability of high-energy lepton colliders to perform precision Higgs measurements. A GNN-based
classifier significantly enhances the signal purity, yielding a signal significance of $\mathscr{Z}\approx 20$ at $5~\mathrm{ab}^{-1}$ for the SM benchmark. A binned likelihood analysis based on observables sensitive to the reconstructed Higgs pair gives projected constraints of $\kappa_\lambda\in[0.76,\,1.31]$ and $\kappa_{2V}\in[0.95,\,1.05]$ at
$95\%$ C.L. for \textit{Case 1} ($\kappa_V=1$). Imposing the SMEFT-motivated relation $\kappa_V=\kappa_{2V}$ weakens the sensitivity to $\kappa_{2V}$, demonstrating the important role of interference effects and assumptions about the underlying Higgs-gauge boson sector. In a broader sense, the dependence of the projected sensitivities on the Higgs-gauge boson couplings confirms that WBF di-Higgs production serves not only as a probe of the Higgs self-interaction but also as an indirect discriminator between underlying SMEFT-like and HEFT-like structures.\\

\begin{acknowledgments}
SB acknowledges ANRF grant CRG/2023/000580.
\end{acknowledgments}
\appendix
\section{Mapping $\kappa$ to EFTs: SMEFT vs. HEFT}
\label{appA} 
The $\kappa$ framework parametrizes NP effects by rescaling SM couplings. A deviation in a given $\kappa$ modifier can originate from: (\textbf{I}) shifts in the SM input parameters, (\textbf{II}) genuine anomalous contributions to the interaction vertex, and (\textbf{III}) additional NP contributions effectively absorbed into the extracted $\kappa$ values. We
discuss below the classes of NP that give rise to each.

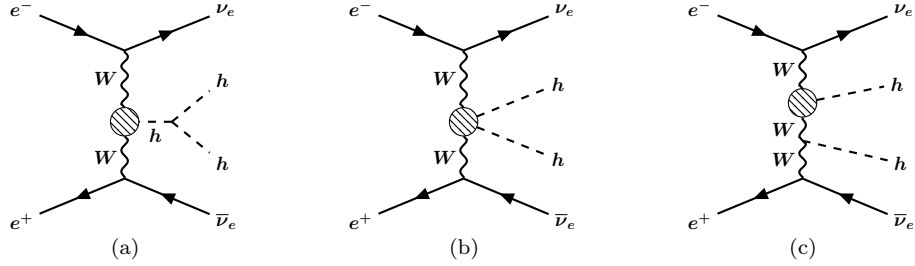
\begin{figure}[htb!]
    \centering
    {\Large
    \subfigure[]{\begin{tikzpicture}[baseline={(current bounding box.center)},style={scale=0.5, transform shape}]
	\begin{feynman}
		\vertex(a);
		\vertex[above left =0.75cm and 2.25cm of a] (a1){$\boldsymbol{e^{-}}$};
        \vertex[blob, below = 1.5cm of a] (c) {};
        \vertex[right = 1.25cm of c] (d);
        \vertex[below = 1.5cm of c] (b);
		\vertex[below left =0.75cm and 2.25cm of b] (b1){$\boldsymbol{e^{+}}$};
        \vertex[above right =0.75cm and 2.25cm of a] (a2){$\boldsymbol{\nu_{e}}$};
        \vertex[below right =0.75cm and 2.25cm of b] (b2){$\boldsymbol{\overline{\nu}_{e}}$};
        \vertex[above right =0.75cm and 1cm of d] (c1){$\boldsymbol{h}$};
        \vertex[below right =0.75cm and 1cm of d] (c2){$\boldsymbol{h}$};
		\diagram*{
			(a1) -- [thick, fermion, arrow size=1.25pt] (a),
                (a) -- [thick, boson, edge label' = $\boldsymbol{W}$, arrow size=1.25pt] (c),
                (b) -- [thick, boson, edge label = $\boldsymbol{W}$, arrow size=1.25pt] (c),
			(b) -- [thick, fermion, arrow size=1.25pt] (b1),
                (a) -- [thick, fermion, arrow size=1.25pt] (a2),
                (b2) -- [thick, fermion, arrow size=1.25pt] (b),
                (d) -- [thick, scalar, edge label = $\boldsymbol{h}$, arrow size=1.25pt] (c),
                (d) -- [thick, scalar, arrow size=1.25pt] (c1),
                (d) -- [thick, scalar, arrow size=1.25pt] (c2)
		};
	\end{feynman}
    \end{tikzpicture}
    \label{fig:2h2v1A}}
    \quad
    \subfigure[]{\begin{tikzpicture}[baseline={(current bounding box.center)},style={scale=0.5, transform shape}]
	\begin{feynman}
		\vertex(a);
		\vertex[above left =0.75cm and 2.25cm of a] (a1){$\boldsymbol{e^{-}}$};
        \vertex[blob, below = 1.5cm of a] (c) {};
        \vertex[below = 1.5cm of c] (b);
		\vertex[below left =0.75cm and 2.25cm of b] (b1){$\boldsymbol{e^{+}}$};
        \vertex[above right =0.75cm and 2.25cm of a] (a2){$\boldsymbol{\nu_{e}}$};
        \vertex[below right =0.75cm and 2.25cm of b] (b2){$\boldsymbol{\overline{\nu}_{e}}$};
        \vertex[above right =1cm and 2.5cm of c] (c1){$\boldsymbol{h}$};
        \vertex[below right =1cm and 2.5cm of c] (c2){$\boldsymbol{h}$};
		\diagram*{
			(a1) -- [thick, fermion, arrow size=1.25pt] (a),
                (a) -- [thick, boson, edge label' = $\boldsymbol{W}$, arrow size=1.25pt] (c),
                (b) -- [thick, boson, edge label = $\boldsymbol{W}$, arrow size=1.25pt] (c),
			(b) -- [thick, fermion, arrow size=1.25pt] (b1),
                (a) -- [thick, fermion, arrow size=1.25pt] (a2),
                (b2) -- [thick, fermion, arrow size=1.25pt] (b),
                (c) -- [thick, scalar, arrow size=1.25pt] (c1),
                (c) -- [thick, scalar, arrow size=1.25pt] (c2)
		};
	\end{feynman}
	\end{tikzpicture}
    \label{fig:2h2v1B}}
    \quad
    \subfigure[]{\begin{tikzpicture}[baseline={(current bounding box.center)},style={scale=0.5, transform shape}]
	\begin{feynman}
		\vertex(a);
		\vertex[above left =0.75cm and 2.25cm of a] (a1){$\boldsymbol{e^{-}}$};
        \vertex[blob, below = 1.0cm of a] (c) {};
        \vertex[below = 1.0cm of c] (d);
        \vertex[below = 1.0cm of d] (b);
		\vertex[below left =0.75cm and 2.25cm of b] (b1){$\boldsymbol{e^{+}}$};
        \vertex[above right =0.75cm and 2.25cm of a] (a2){$\boldsymbol{\nu_{e}}$};
        \vertex[below right =0.75cm and 2.25cm of b] (b2){$\boldsymbol{\overline{\nu}_{e}}$};
        \vertex[above right =0.5cm and 2.5cm of c] (c1){$\boldsymbol{h}$};
        \vertex[below right =0.25cm and 2.25cm of d] (c2){$\boldsymbol{h}$};
		\diagram*{
			(a1) -- [thick, fermion, arrow size=1.25pt] (a),
                (a) -- [thick, boson, edge label' = $\boldsymbol{W}$, arrow size=1.25pt] (c),
                (c) -- [thick, boson, edge label' = $\boldsymbol{W}$, arrow size=1.25pt] (d),
                (b) -- [thick, boson, edge label = $\boldsymbol{W}$, arrow size=1.25pt] (d),
			(b) -- [thick, fermion, arrow size=1.25pt] (b1),
                (a) -- [thick, fermion, arrow size=1.25pt] (a2),
                (b2) -- [thick, fermion, arrow size=1.25pt] (b),
                (c) -- [thick, scalar, arrow size=1.25pt] (c1),
                (d) -- [thick, scalar, arrow size=1.25pt] (c2)
		};
	\end{feynman}
	\end{tikzpicture}
    \label{fig:2h2v1C}}
    \quad
    \subfigure[]{\begin{tikzpicture}[baseline={(current bounding box.center)},style={scale=0.5, transform shape}]
	\begin{feynman}
		\vertex(a);
		\vertex[above left =0.75cm and 2.25cm of a] (a1){$\boldsymbol{e^{-}}$};
        \vertex[blob, below = 1.0cm of a] (c) {};
        \vertex[below = 1.0cm of c] (d);
        \vertex[below = 1.0cm of d] (b);
		\vertex[below left =0.75cm and 2.25cm of b] (b1){$\boldsymbol{e^{+}}$};
        \vertex[above right =0.75cm and 2.25cm of a] (a2){$\boldsymbol{\nu_{e}}$};
        \vertex[below right =0.75cm and 2.25cm of b] (b2){$\boldsymbol{\overline{\nu}_{e}}$};
        \vertex[above right =0.5cm and 2.5cm of c] (c1){$\boldsymbol{h}$};
        \vertex[below right =0.25cm and 2.25cm of d] (c2){$\boldsymbol{h}$};
		\diagram*{
			(a1) -- [thick, fermion, arrow size=1.25pt] (a),
                (a) -- [thick, boson, edge label' = $\boldsymbol{W}$, arrow size=1.25pt] (c),
                (c) -- [thick, boson, edge label' = $\boldsymbol{W}$, arrow size=1.25pt] (d),
                (b) -- [thick, boson, edge label = $\boldsymbol{W}$, arrow size=1.25pt] (d),
			(b) -- [thick, fermion, arrow size=1.25pt] (b1),
                (a) -- [thick, fermion, arrow size=1.25pt] (a2),
                (b2) -- [thick, fermion, arrow size=1.25pt] (b),
                (c) -- [thick, scalar, arrow size=1.25pt] (c2),
                (d) -- [thick, scalar, arrow size=1.25pt] (c1)
		};
	\end{feynman}
	\end{tikzpicture}
    \label{fig:2h2v1D}}}
    \caption{Feynman diagrams for EFT modifications to the $hWW/hhWW$
    vertex in di-Higgs production via WBF.}
     \label{fig:dihiggs1} 
\end{figure}

\paragraph{Modification to $hWW/hhWW$ vertices.}
The different ways in which the $hWW$ and $hhWW$ interactions can be modified in the WBF process are illustrated in Fig.~\ref{fig:dihiggs1}. In the SM, these interactions originate from the Higgs kinetic term,
\begin{equation}
\mathcal{L}^{hWW}_{\rm SM} \supset
\underbrace{\frac{2m_W^2}{v}}_{g^{hWW}_{\rm SM}} h\,W_\mu^+W^{-\mu} +
\underbrace{\frac{m_W^2}{v^2}}_{g^{hhWW}_{\rm SM}} h^2\,W_\mu^+W^{-\mu}.
\end{equation}
Since both couplings arise from the same gauge-invariant operator, they are intrinsically correlated in the SM and receive simultaneous modifications in any extension preserving the linearly realized $SU(2)_L\times U(1)_Y$ symmetry. In SMEFT, corrections are generated by
higher-dimensional operators like,
\begin{equation}
\mathcal{O}_{HW}=(H^\dagger H)\,W^i_{\mu\nu}W^{i\mu\nu},
\end{equation}
which induces momentum-dependent corrections to both vertices with Lorentz structures different from the SM. To a good approximation, such effects lead to the correlated relation $\kappa_V=\kappa_{2V}$, corresponding to \textit{Case 2} of the main text.

In theories based on a non-linear realization of EWSB, such as HEFT, the $hWW$ and $hhWW$ interactions need not originate from the same operator and can vary independently. Operators of the form
\begin{equation}
h\,W^+_{\mu\nu}W^{-\mu\nu}, \qquad h^2\,W^+_{\mu\nu}W^{-\mu\nu},
\end{equation}
appear with independent coefficients, leading naturally to uncorrelated modifications of $\kappa_V$ and $\kappa_{2V}$, corresponding to \textit{Case 1}. The two scenarios therefore represent qualitatively distinct realizations of EWSB, and the sensitivity of WBF di-Higgs
production to $\kappa_V$ and $\kappa_{2V}$ can provide indirect information about the underlying structure of NP.\\

\begin{figure}[htb!]
    \centering
    {\Large
    \subfigure[]{\begin{tikzpicture}[baseline={(current bounding box.center)},style={scale=0.5, transform shape}]
	\begin{feynman}
		\vertex[blob](c){};
		\vertex[above left =1.25cm and 2.75cm of c] (a1){$\boldsymbol{e^{-}}$};
        \vertex[ below right = 1.5cm and 1.25 of c] (d);
        \vertex[below = 3cm of c] (b);
		\vertex[below left =0.75cm and 2.25cm of b] (b1){$\boldsymbol{e^{+}}$};
        \vertex[above right =1.25cm and 2.75cm of c] (a2){$\boldsymbol{\nu_{e}}$};
        \vertex[below right =0.75cm and 2.25cm of b] (b2){$\boldsymbol{\overline{\nu}_{e}}$};
        \vertex[above right =0.75cm and 1cm of d] (c1){$\boldsymbol{h}$};
        \vertex[below right =0.75cm and 1cm of d] (c2){$\boldsymbol{h}$};
		\diagram*{
			(a1) -- [thick, fermion, arrow size=1.25pt] (c),
                (b) -- [thick, boson, edge label = $\boldsymbol{W}$, arrow size=1.25pt] (c),
			(b) -- [thick, fermion, arrow size=1.25pt] (b1),
                (c) -- [thick, fermion, arrow size=1.25pt] (a2),
                (b2) -- [thick, fermion, arrow size=1.25pt] (b),
                (d) -- [thick, scalar, edge label' = $\boldsymbol{h}$, arrow size=1.25pt] (c),
                (d) -- [thick, scalar, arrow size=1.25pt] (c1),
                (d) -- [thick, scalar, arrow size=1.25pt] (c2)
		};
	\end{feynman}
    \end{tikzpicture}
    \label{fig:evwhA}}
    \quad
    \subfigure[]{\begin{tikzpicture}[baseline={(current bounding box.center)},style={scale=0.5, transform shape}]
	\begin{feynman}
		\vertex[blob] (c) {};
		\vertex[above left =1.25cm and 2.75cm of c] (a1){$\boldsymbol{e^{-}}$};
        \vertex[below = 3cm of c] (b);
		\vertex[below left =0.75cm and 2.25cm of b] (b1){$\boldsymbol{e^{+}}$};
        \vertex[above right =1.25cm and 2.75cm of c] (a2){$\boldsymbol{\nu_{e}}$};
        \vertex[below right =0.75cm and 2.25cm of b] (b2){$\boldsymbol{\overline{\nu}_{e}}$};
        \vertex[below right =0.75cm and 2.5cm of c] (c1){$\boldsymbol{h}$};
        \vertex[below right =2.25cm and 2.5cm of c] (c2){$\boldsymbol{h}$};
		\diagram*{
			(a1) -- [thick, fermion, arrow size=1.25pt] (c),
                (b) -- [thick, boson, edge label = $\boldsymbol{W}$, arrow size=1.25pt] (c),
			(b) -- [thick, fermion, arrow size=1.25pt] (b1),
                (c) -- [thick, fermion, arrow size=1.25pt] (a2),
                (b2) -- [thick, fermion, arrow size=1.25pt] (b),
                (c) -- [thick, scalar, arrow size=1.25pt] (c1),
                (c) -- [thick, scalar, arrow size=1.25pt] (c2)
		};
	\end{feynman}
	\end{tikzpicture}
        \label{fig:evwhB}}
    \quad
    \subfigure[]{\begin{tikzpicture}[baseline={(current bounding box.center)},style={scale=0.5, transform shape}]
	\begin{feynman}
		\vertex[blob] (c) {};
		\vertex[above left =1.25cm and 2.75cm of c] (a1){$\boldsymbol{e^{-}}$};
        \vertex[below = 2.0cm of c] (d);
        \vertex[below = 1.0cm of d] (b);
		\vertex[below left =0.75cm and 2.25cm of b] (b1){$\boldsymbol{e^{+}}$};
        \vertex[above right =1.25cm and 2.75cm of c] (a2){$\boldsymbol{\nu_{e}}$};
        \vertex[below right =0.75cm and 2.25cm of b] (b2){$\boldsymbol{\overline{\nu}_{e}}$};
        \vertex[below right =0.5cm and 2.5cm of c] (c1){$\boldsymbol{h}$};
        \vertex[below right =0.25cm and 2.25cm of d] (c2){$\boldsymbol{h}$};
		\diagram*{
			(a1) -- [thick, fermion, arrow size=1.25pt] (c),
                (c) -- [thick, boson, edge label' = $\boldsymbol{W}$, arrow size=1.25pt] (d),
                (b) -- [thick, boson, edge label = $\boldsymbol{W}$, arrow size=1.25pt] (d),
			(b) -- [thick, fermion, arrow size=1.25pt] (b1),
                (c) -- [thick, fermion, arrow size=1.25pt] (a2),
                (b2) -- [thick, fermion, arrow size=1.25pt] (b),
                (c) -- [thick, scalar, arrow size=1.25pt] (c1),
                (d) -- [thick, scalar, arrow size=1.25pt] (c2)
		};
	\end{feynman}
	\end{tikzpicture}
    \label{fig:evwhC}}
    \quad
    \subfigure[]{\begin{tikzpicture}[baseline={(current bounding box.center)},style={scale=0.5, transform shape}]
	\begin{feynman}
		\vertex[blob] (c) {};
		\vertex[above left =1.25cm and 2.75cm of c] (a1){$\boldsymbol{e^{-}}$};
        \vertex[below = 2.0cm of c] (d);
        \vertex[below = 1.0cm of d] (b);
		\vertex[below left =0.75cm and 2.25cm of b] (b1){$\boldsymbol{e^{+}}$};
        \vertex[above right =1.25cm and 2.75cm of c] (a2){$\boldsymbol{\nu_{e}}$};
        \vertex[below right =0.75cm and 2.25cm of b] (b2){$\boldsymbol{\overline{\nu}_{e}}$};
        \vertex[below right =0.5cm and 2.5cm of c] (c1){$\boldsymbol{h}$};
        \vertex[below right =0.25cm and 2.25cm of d] (c2){$\boldsymbol{h}$};
		\diagram*{
			(a1) -- [thick, fermion, arrow size=1.25pt] (c),
                (c) -- [thick, boson, edge label' = $\boldsymbol{W}$, arrow size=1.25pt] (d),
                (b) -- [thick, boson, edge label = $\boldsymbol{W}$, arrow size=1.25pt] (d),
			(b) -- [thick, fermion, arrow size=1.25pt] (b1),
                (c) -- [thick, fermion, arrow size=1.25pt] (a2),
                (b2) -- [thick, fermion, arrow size=1.25pt] (b),
                (c) -- [thick, scalar, arrow size=1.25pt] (c2),
                (d) -- [thick, scalar, arrow size=1.25pt] (c1)
		};
	\end{feynman}
	\end{tikzpicture}
    \label{fig:evwhD}}}
    \caption{ Feynman diagrams for EFT-induced $e\nu Wh/e\nu Whh$ contact
    vertices in di-Higgs production via WBF.}
    \label{fig:evwh}
\end{figure}
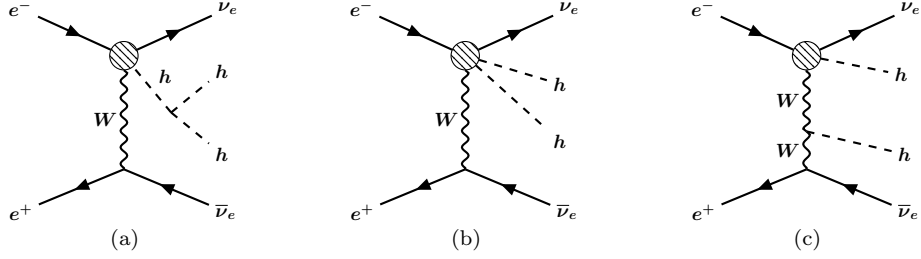

\paragraph{Introduction of $e\nu Wh/e\nu Whh$ vertices.}
EFTs can also induce contact interactions involving $e\nu Wh$ and $e\nu Whh$ vertices, as illustrated in Fig.~\ref{fig:evwh}. Although these do not directly modify the $hWW$ or $hhWW$ vertices, they alter the production amplitude and, if unaccounted for, can be absorbed into the extracted $\kappa$ parameters. Within SMEFT, such contributions arise from
Higgs-current operators (e.g.\ $\mathcal{O}^{(3)}_{H\ell}=(\bar{\ell}\tau^a\gamma^\mu\ell)(H^\dagger\tau^a i\overleftrightarrow{D_\mu}H)$) and dipole operators (e.g.\ $\mathcal{O}_{eW}=(\bar{\ell}\tau^a\sigma^{\mu\nu}e)HW^a_{\mu\nu}$). Under the assumption of flavor universality, however, these operators are tightly constrained by electroweak precision observables and low-energy measurements. A consistent matching onto the $\kappa$ framework requires a more complete EFT treatment and lies beyond the scope of the present study.\\

\paragraph{Modification to the $hhh$ self-coupling.}
Modifications of the trilinear Higgs self-coupling can arise not only from a rescaling of the SM $hhh$ vertex, as generated in SMEFT by the operator $\mathcal{O}_{H}=(H^\dagger H)^3$, but also from derivative self-interactions through operators such as $\mathcal{O}_{H\Box}=(H^\dagger H)\Box(H^\dagger H)$, which introduce momentum-dependent modifications to the $hhh$ vertex and alter the kinematic distributions of di-Higgs production in addition to
the overall rate. In the simplified analysis presented here, these effects are absorbed into an effective modifier $\kappa_\lambda$; a complete treatment within a full EFT basis is left for future work.

\section{WBF Production Cross Section}
\label{appB}
\begin{figure*}[htb!]
    \centering
    \includegraphics[width=0.9\textwidth]{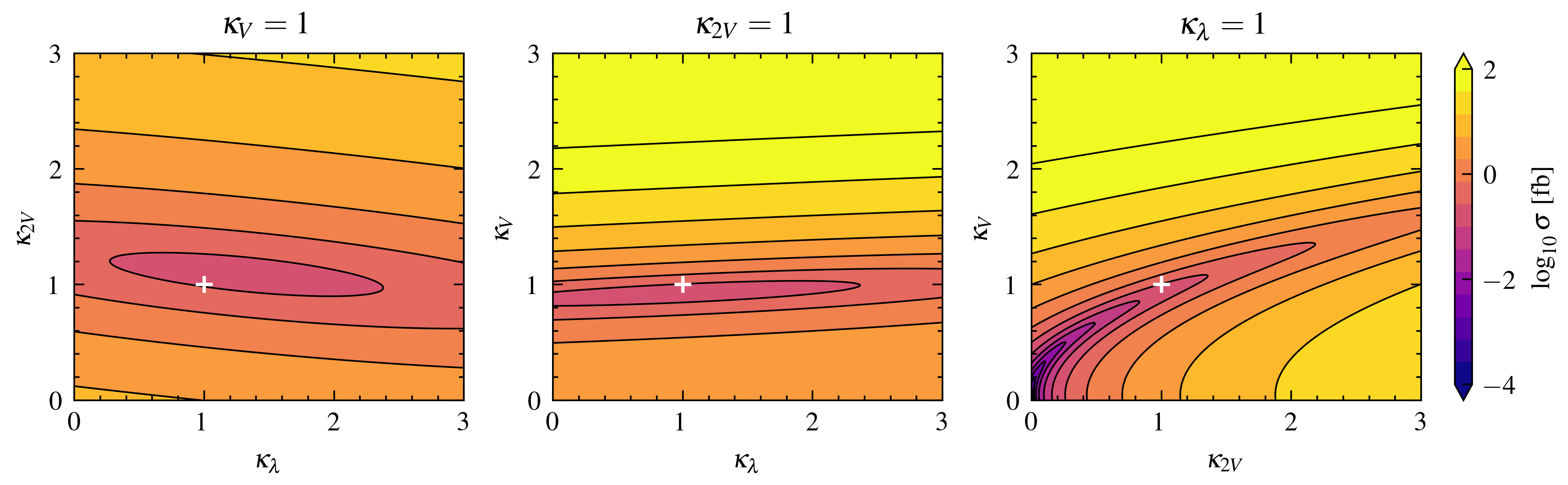}
    \caption{WBF di-Higgs production cross sections in different parameter
    planes: $\kappa_{\lambda}$-$\kappa_{2V}$ with $\kappa_{V}=1$
    (\textit{left}); $\kappa_{\lambda}$-$\kappa_{V}$ with $\kappa_{2V}=1$
    (\textit{center}); $\kappa_{2V}$-$\kappa_{V}$ with $\kappa_{\lambda}=1$
    (\textit{right}). The symbol $+$ denotes the SM prediction
    $\kappa_{\lambda}=\kappa_{2V}=\kappa_{V}=1$.}
    \label{fig:xsk}
\end{figure*}

The WBF di-Higgs production cross section can be parameterized in terms of the coupling modifiers as 
\begin{equation}
\begin{split}
    & \sigma(\kappa_{\lambda},\kappa_{2V},\kappa_{V}) =
    \kappa_{\lambda}^{2}\kappa_{V}^{2}C_{11}
    + \kappa_{2V}^{2}C_{22}
    + \kappa_{V}^{4}C_{33} \\ & + \kappa_{\lambda}\kappa_{2V}\kappa_{V}C_{12}
    + \kappa_{\lambda}\kappa_{V}^{3}C_{13}
    + \kappa_{2V}\kappa_{V}^{2}C_{23}\,,
\end{split}
\end{equation}
where the coefficients $C_{ij}$ encode the individual and interference contributions of the underlying amplitudes. In Fig.~\ref{fig:xsk}, we present the WBF di-Higgs production cross section in the $\kappa_{\lambda}$-$\kappa_{2V}$ plane with $\kappa_{V}=1$ (\textit{left}),
the $\kappa_{\lambda}$-$\kappa_{V}$ plane with $\kappa_{2V}=1$ (\textit{center}), and the $\kappa_{2V}$-$\kappa_{V}$ plane with $\kappa_{\lambda}=1$ (\textit{right}). The symbol `$+$' denotes the SM prediction.

\section{Graph Neural Network Event Classifier}
\label{appC}

We employ a heterogeneous graph neural network (GNN) to discriminate the di-Higgs signal $hh\slashed{E}$ from two principal backgrounds: single-Higgs associated production $hb\bar{b}\slashed{E}$ and top-quark processes ($t\bar{t}$, $t\bar{t}Z$, $t\bar{t}h$, collectively $t\bar{t}X$). The QCD-induced $4b\slashed{E}$ background is rendered negligible by the Higgs mass-window selection described in the main text and is dropped from the analysis. All samples are generated at leading order and passed through the same selection described in the main text; the resulting efficiencies and expected yields are collected in Table~\ref{tab:event}. After kinematic selection alone, the significance at $1~\mathrm{ab}^{-1}$ is $\mathscr{Z}\approx 3$, motivating the GNN-based classifier.

\begin{table*}[htb!]
    \centering
    \caption{Post-selection efficiency $\epsilon$ and expected event yields
    $N$ at integrated luminosities of 1, 3, and 5~ab$^{-1}$, computed from
    $10^6$ parton-level events per process passed through the full
    detector simulation chain.}
    \label{tab:event}
    \renewcommand{\arraystretch}{1.3}
    \begin{tabular*}{0.75\textwidth}{@{\extracolsep{\fill}} c c c c @{}}
    \hline
    Process & $\sigma$ [fb] & $\epsilon$ &
    $N\;(1,3,5~\mathrm{ab}^{-1})$ \\
    \hline
    $hh\slashed{E}$         & $0.845$ & $0.032$ & $(27.0,\;81.1,\;135.2)$ \\
    $hb\bar{b}\slashed{E}$  & $1.509$ & $0.002$ & $(3.0,\;9.1,\;15.2)$ \\
    $t\bar{t}X$             & $21.19$ & $0.003$ & $(63.6,\;190.7,\;317.9)$ \\
    $4b\slashed{E}$         & $0.656$ & $\mathcal{O}(10^{-6})$ & $-$ \\
    \hline
    \end{tabular*}
\end{table*}

Each event is encoded as a heterogeneous directed graph $\mathcal{G}=(\mathcal{V},\mathcal{E})$ with two node types. Every reconstructed jet defines a node carrying a 14-dimensional feature vector,
\begin{equation}
\begin{split}
\mathbf{x}^{(\text{jet})} &=
  \bigl(p_T,\,\phi,\,\eta,\,m,\,N_{\text{ch}},\, N_{\mathrm{ne}},\,E_{\text{had}}/E_{\text{em}},\\ &\qquad\,b_{\text{tag}},\,\text{flav},\,T,\,\Delta\eta,\,\Delta\phi,\,f_{\text{ch}},\,f_{\text{neu}}\bigr),
\end{split}
\end{equation}
where $p_T$, $\phi$, $\eta$, and $m$ are the standard kinematic variables, $N_{\text{ch}}(N_{\text{ne}})$ is the charged (neutral)-particle multiplicity, $E_{\text{had}}/E_{\text{em}}$ is the hadronic-to-electromagnetic energy ratio; $b_{\text{tag}}$ and $\text{flav}$ encode flavour information; $T$
is the particle time of flight, $(\Delta\eta,\Delta\phi)$ parametrize the jet radius; and $(f_{\text{ch}},f_{\text{neu}})$ are the charged and neutral energy fractions. The missing transverse energy is represented by a single node per event,
\begin{equation}
  \mathbf{x}^{(\text{met})} = \bigl(\slashed{E}_T,\,\phi_{\slashed{E}_T}\bigr),
\end{equation}
which carries no edges and its information enters the network exclusively through the node-feature projection pathway.

Jet nodes are normalized per-event via a $z$-score,
\begin{equation}
  \tilde{x}_{ij} = \frac{x_{ij} - \mu_j}{\sigma_j + \varepsilon},
  \qquad \varepsilon = 10^{-8},
\end{equation}
while the single MET node bypasses per-event normalization (degenerate for a singleton) and is normalized by downstream batch normalization layers.

A directed edge is drawn between jet nodes $i$ and $j$ whenever $\Delta R_{ij}\le 1.5$, with edge attribute
\begin{equation}
  \mathbf{e}_{ij} = \bigl(\Delta\eta_{ij},\;\Delta\phi_{ij},\;\Delta R_{ij}\bigr),
\end{equation}
where $\Delta\phi_{ij}$ is wrapped to $(-\pi,\pi]$. This locality criterion reflects the calorimeter clustering scale and ensures that message passing propagates correlated energy deposits while suppressing long-range spurious connections.

The backbone is a three-layer GATv2~\cite{brody2022} graph attention network, which resolves the static-attention limitation of the original GAT~\cite{veligat} by making the attention score jointly dependent on both endpoint representations. For a directed edge $j\to i$ the attention score is
\begin{equation}
  e(\mathbf{h}_i,\mathbf{h}_j)
  = \mathbf{a}^\top\,\mathrm{LeakyReLU}
    \!\left(\mathbf{W}\!\left[\mathbf{h}_i\,\|\,\mathbf{h}_j\right]\right),
\end{equation}
normalized via softmax over the neighborhood $\mathcal{N}(i)$,
\begin{equation}
  \alpha_{ij}
  = \frac{\exp\!\left(e(\mathbf{h}_i,\mathbf{h}_j)\right)}
         {\sum_{j'\in\mathcal{N}(i)}
          \exp\!\left(e(\mathbf{h}_i,\mathbf{h}_{j'})\right)},
\end{equation}
to yield the aggregated node update
\begin{equation}
  \mathbf{h}'_i
  = \sigma\!\left(\sum_{j\in\mathcal{N}(i)}\alpha_{ij}\,\mathbf{W}\mathbf{h}_j\right).
\end{equation}
Input jet and MET features occupy spaces of dimensionality 13 and 2, respectively, and are first mapped to a common $d=128$-dimensional latent space through separate encoders, each consisting of a linear projection, LayerNorm~\cite{layernorm}, ELU activation~\cite{elu}, and
dropout~\cite{dropout}, so that both modalities enter the message-passing layers on a comparable footing irrespective of their raw feature scales. Three successive GATv2 layers then process the jet graph (with 4, 2, and 2 attention heads, progressively compressing to $d/2=64$), while MET is updated through parallel linear projections at each layer. After message
passing, the jet graph is pooled via mean, sum, and global-attention pooling; the MET embedding is mean-pooled. The four pooled vectors are concatenated into a $4\times 64=256$-dimensional event representation and passed to a four-layer MLP classifier ($256\to128\to64\to32\to3$) with
LayerNorm and ELU activations at each hidden layer. The total trainable parameter count is $\approx2.6\times10^5$.

Class imbalance is addressed by assigning each class $c$ the weight $w_c = \frac{N_{\text{total}}}{C\cdot N_c}$, where $N_{\text{total}}$ is the total number of training events, $C=3$ is the number of classes, and $N_c$ is the number of events in class $c$,
following the balanced prescription of \textsc{scikit-learn}~\cite{scikit-learn}. The weighted cross-entropy loss
\begin{equation}
  \mathcal{L}
  = -\frac{1}{|\mathcal{B}|}\sum_{(G,y)\in\mathcal{B}}
    w_y\sum_{c=0}^{C-1}\mathbf{1}[y=c]\,\log\hat{p}_c,
\end{equation}
where $\hat{p}_c=\mathrm{softmax}(\hat{\mathbf{y}})_c$ and $\mathcal{B}$ is a mini-batch of 128 graphs, is minimized using \textsc{Adam}~\cite{adam} with initial learning rate $10^{-4}$ and $L_2$ weight decay $10^{-5}$. Gradients are clipped at unit norm. The learning rate follows cosine annealing~\cite{cosine},
\begin{equation}
  \eta_t = \eta_{\min}
    + \tfrac{1}{2}(\eta_{\max}-\eta_{\min})
      \left(1+\cos\frac{T_{\mathrm{cur}}\pi}{T_{\mathrm{max}}}\right),
\end{equation}
with $T_{\mathrm{max}}=100$ epochs per cycle, and a dropout rate of 0.2 is applied at all hidden layers. Training employed an 80/20 train/test split and converged in approximately 2 hours on a standard CPU.

\begin{figure}[htb!]
    \centering
    \includegraphics[width=0.75\columnwidth]{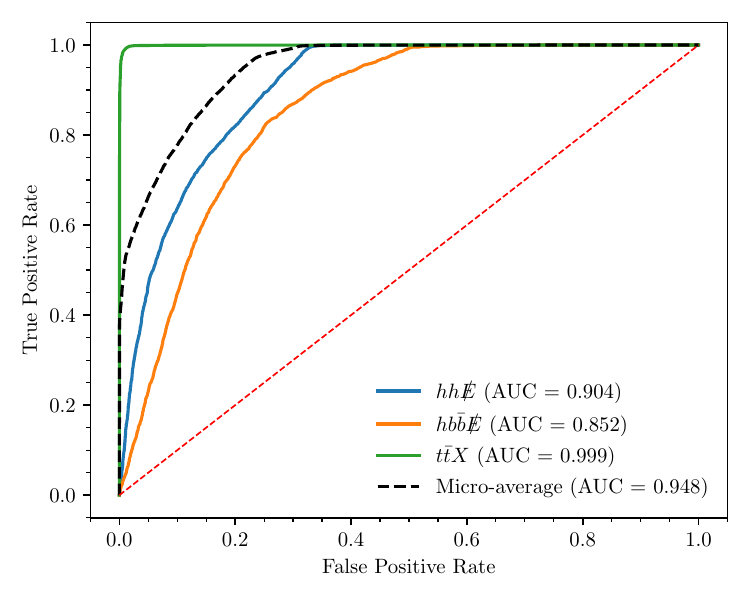} \\
    \includegraphics[width=0.75\columnwidth]{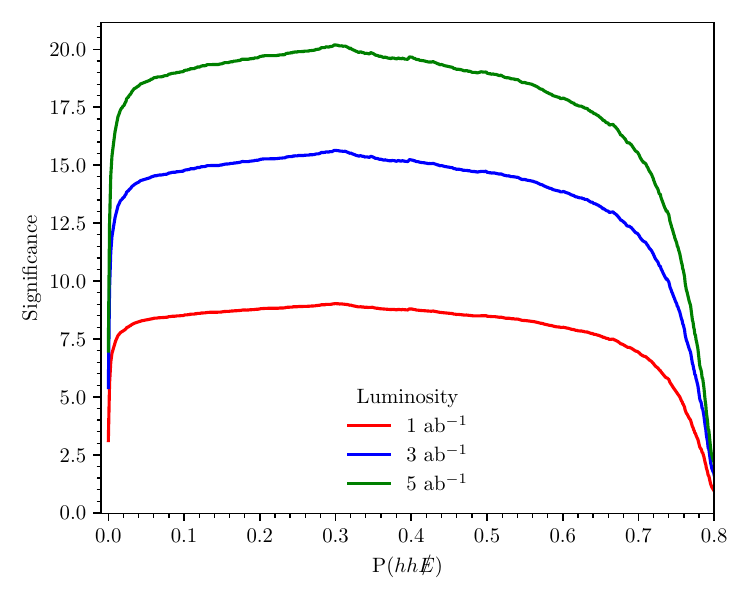}
    \caption{Left: One-vs-rest ROC curves for the three-class GATv2 classifier. Right: Profile-likelihood significance $\mathscr{Z}$ as a function of signal-score threshold at integrated luminosities of 1, 3, and 5~ab$^{-1}$.}
    \label{fig:clicGNN}
\end{figure}

Fig.~\ref{fig:clicGNN} presents the receiver operating characteristic (ROC) curves and the profile-likelihood significance of Eq.~\eqref{eq:z} as a function of signal-score threshold, evaluated at integrated luminosities of 1, 3, and 5~ab$^{-1}$.

The classifier achieves one-vs-rest AUC values of 0.90 ($hh\slashed{E}$), 0.85 ($hb\bar{b}\slashed{E}$), and 0.99 ($t\bar{t}X$), with a micro-averaged AUC of 0.95. The near-perfect separation of the top-quark background reflects its distinctive high-multiplicity, high-$p_T$ topology, which is well-resolved by the flavour-aware graph structure. The $hb\bar{b}\slashed{E}$ background presents the greatest challenge owing to its kinematic similarity to the signal; nonetheless, the attention mechanism could achieve an AUC above 0.85.

The significance panel demonstrates that the result is well above the $5\sigma$ discovery threshold already at $1~\mathrm{ab}^{-1}$ for signal-score thresholds in the range $P(hh\slashed{E})\sim 0.3$--$0.4$. At luminosities of 3 and 5~ab$^{-1}$ the significance grows
substantially, reaching $\mathscr{Z}\sim 15$ and $\mathscr{Z}\sim 20$, respectively, in the same threshold region, demonstrating that the GNN classifier sustains robust signal extraction well into the high-luminosity CLIC program. The broad plateau in $\mathscr{Z}$ around the optimal threshold indicates that the result is stable against the precise choice of working point.

\section{WBF di-Higgs Signal Significance}\label{appD}
The overall selection and classifier efficiencies for different parameter benchmarks for the signal,
$\epsilon_{S}(\kappa_{\lambda},\kappa_{2V},\kappa_{V})$, are:
\begin{equation}
\begin{split}
    \epsilon_{S}(2,1,1)&:0.03745\,, \\
    \epsilon_{S}(1,2,1)&:0.01673\,, \\
    \epsilon_{S}(1,1,2)&:0.02390\,, \\
    \epsilon_{S}(3,1,1)&:0.03423\,, \\
    \epsilon_{S}(1,3,1)&:0.01778\,, \\
    \epsilon_{S}(1,1,3)&:0.02423\,, \\
    \epsilon_{S}(2,2,1)&:0.01981\,, \\
    \epsilon_{S}(1,2,2)&:0.02647\,, \\
    \epsilon_{S}(2,1,2)&:0.02328 \,.
\end{split}
\end{equation}
The background efficiency is sensitive to $\kappa_V$; for two values away from the SM: $\epsilon_{B}(\kappa_V=2)=0.00077$ and $\epsilon_{B}(\kappa_V=3)=0.00086$. The signal efficiencies exhibit a mild dependence on the anomalous couplings, with a slight increase for
larger $\kappa_\lambda$, while variations in $\kappa_{2V}$ and $\kappa_V$ have a comparatively smaller impact, indicating that the overall sensitivity is predominantly driven by changes in the production cross section and the underlying interference pattern.

\begin{figure*}[htb!]
    \centering
    \includegraphics[width=0.9\textwidth]{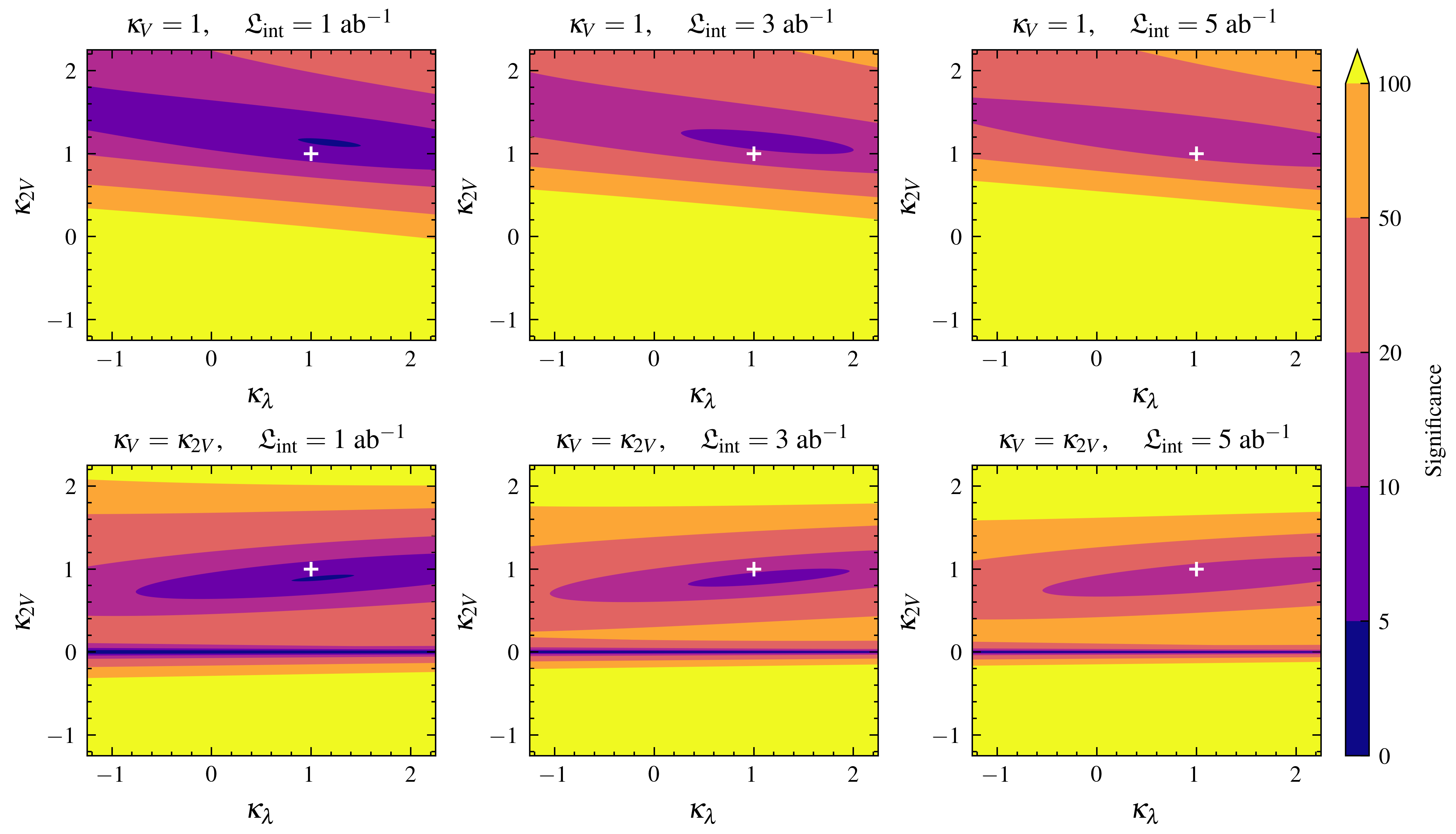}
    \caption{Signal significance in different parameter planes:
    $\kappa_{\lambda}$-$\kappa_{2V}$ with $\kappa_V=1$ (\textit{left});
    $\kappa_{\lambda}$-$\kappa_V$ with $\kappa_{2V}=1$ (\textit{center});
    $\kappa_{2V}$-$\kappa_V$ with $\kappa_{\lambda}=1$ (\textit{right}).
    The symbol $+$ denotes the SM point
    $\kappa_{\lambda}=\kappa_{2V}=\kappa_V=1$. The \textit{top} panel
    corresponds to \textit{Case 1} ($\kappa_V=1$) and the \textit{bottom}
    panel to \textit{Case 2} ($\kappa_V=\kappa_{2V}$).}
    \label{fig:sign}
\end{figure*}

In Fig.~\ref{fig:sign}, we present the signal significance in the $\kappa_{\lambda}$-$\kappa_{2V}$ ($\kappa_V=1$, \textit{left}), $\kappa_{\lambda}$-$\kappa_V$ ($\kappa_{2V}=1$, \textit{center}), and $\kappa_{2V}$-$\kappa_V$ ($\kappa_{\lambda}=1$, \textit{right}) planes.
The significance closely follows the behaviour of the corresponding production cross sections, exhibiting regions of enhancement and suppression from amplitude interference. In \textit{Case 1} ($\kappa_V=1$, top panel), the significance shows a strong dependence on $\kappa_{2V}$,
traceable to the strong interference between the amplitudes proportional to $\kappa_{2V}$ and $\kappa_V^2$. In \textit{Case 2} ($\kappa_V=\kappa_{2V}$, bottom panel), the significance contours differ appreciably in both shape and location of maximal sensitivity, reflecting the modified interference pattern from the coupling correlation. The SM benchmark point ($+$ marker)
lies in a region of substantial significance in both scenarios, highlighting the excellent potential of the 3~TeV CLIC to probe the Higgs self-coupling through the WBF di-Higgs channel.

\section{Effect of Luminosity and Systematics on Sensitivity}
\label{appE}
To illustrate the luminosity dependence of the projected sensitivities, Table~\ref{tab:lum1} presents the 1D limits on $\kappa_{\lambda}$ and $\kappa_{2V}$ at $\mathfrak{L}_{\rm int}=1$, 3, and 5~ab$^{-1}$. As expected, the sensitivity improves with luminosity owing to larger signal
statistics.

\begin{table*}[htb!]
    \centering
    \caption{1D sensitivity limits at $\mathfrak{L}_{\rm int}=1$, 3, and
    5~ab$^{-1}$.}
    \label{tab:lum1}
    \renewcommand{\arraystretch}{1.3}
    \begin{tabular*}{0.75\textwidth}{@{\extracolsep{\fill}} c c c c c @{}}
    \hline
    $\mathfrak{L}_{\rm int}$ & C.L. & $\kappa_{\lambda}$ &
    $\kappa_{2V}$ (\textit{Case 1}) & $\kappa_{2V}$ (\textit{Case 2}) \\
    \hline
    \multirow{2}*{1 ab$^{-1}$} & $68\%$ & $[0.74,1.36]$ & $[0.95,1.06]$ & $[0.88,1.11]$ \\
     & $95\%$ & $[0.52,2.10]$ & $[0.90,1.12]$ & $[0.42,1.14]$ \\ \hline
    \multirow{2}*{3 ab$^{-1}$} & $68\%$ & $[0.84,1.19]$ & $[0.97,1.03]$ & $[0.93,1.06]$ \\
     & $95\%$ & $[0.71,1.43]$ & $[0.94,1.06]$ & $[0.86,1.12]$ \\ \hline
    \multirow{2}*{5 ab$^{-1}$} & $68\%$ & $[0.88,1.14]$ & $[0.98,1.02]$ & $[0.95,1.05]$ \\
     & $95\%$ & $[0.76,1.31]$ & $[0.95,1.05]$ & $[0.90,1.10]$ \\
    \hline
    \end{tabular*}
\end{table*}

\begin{figure}[htb!]
    \centering
    \includegraphics[width=0.75\columnwidth]{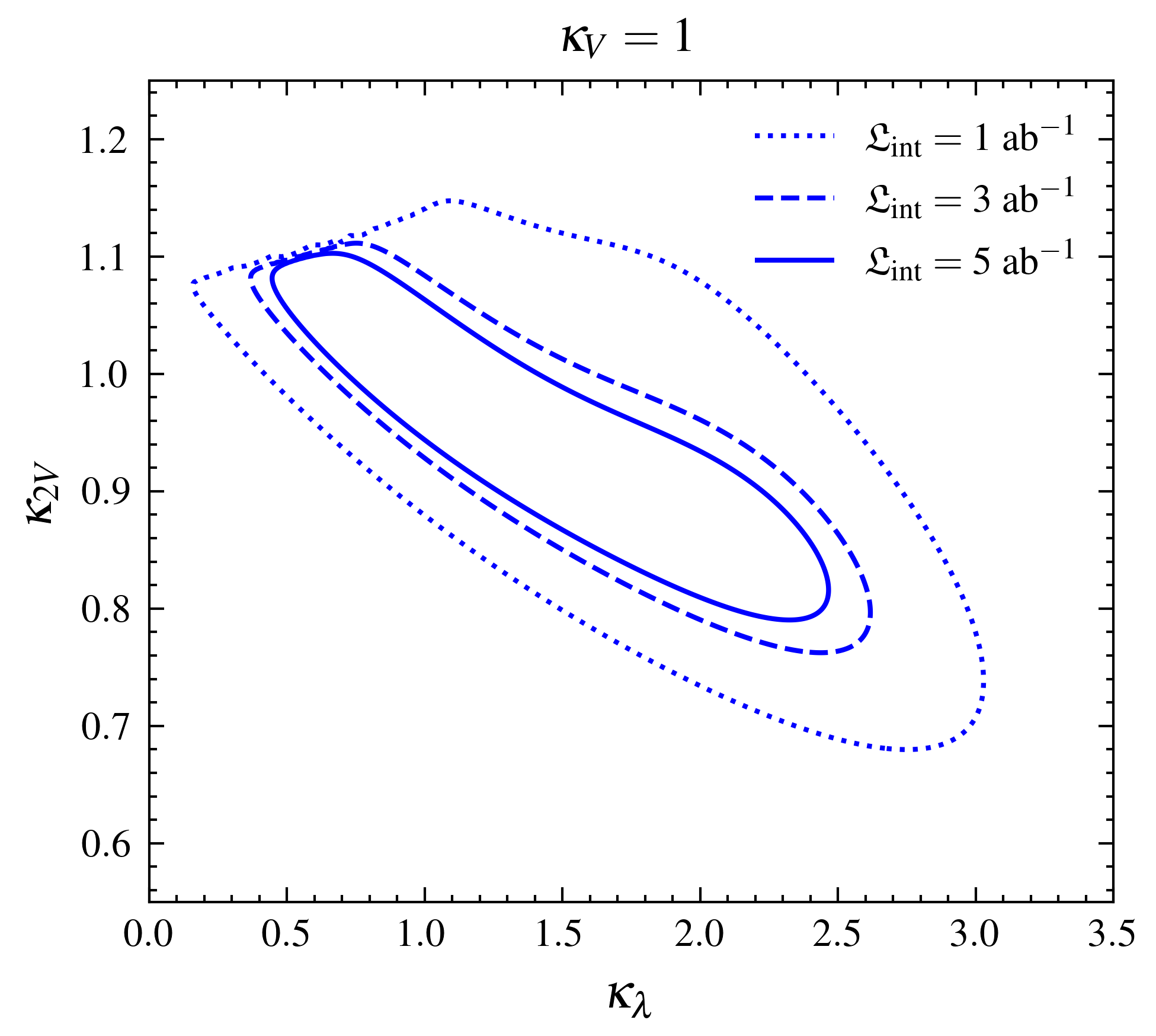}
    \includegraphics[width=0.75\columnwidth]{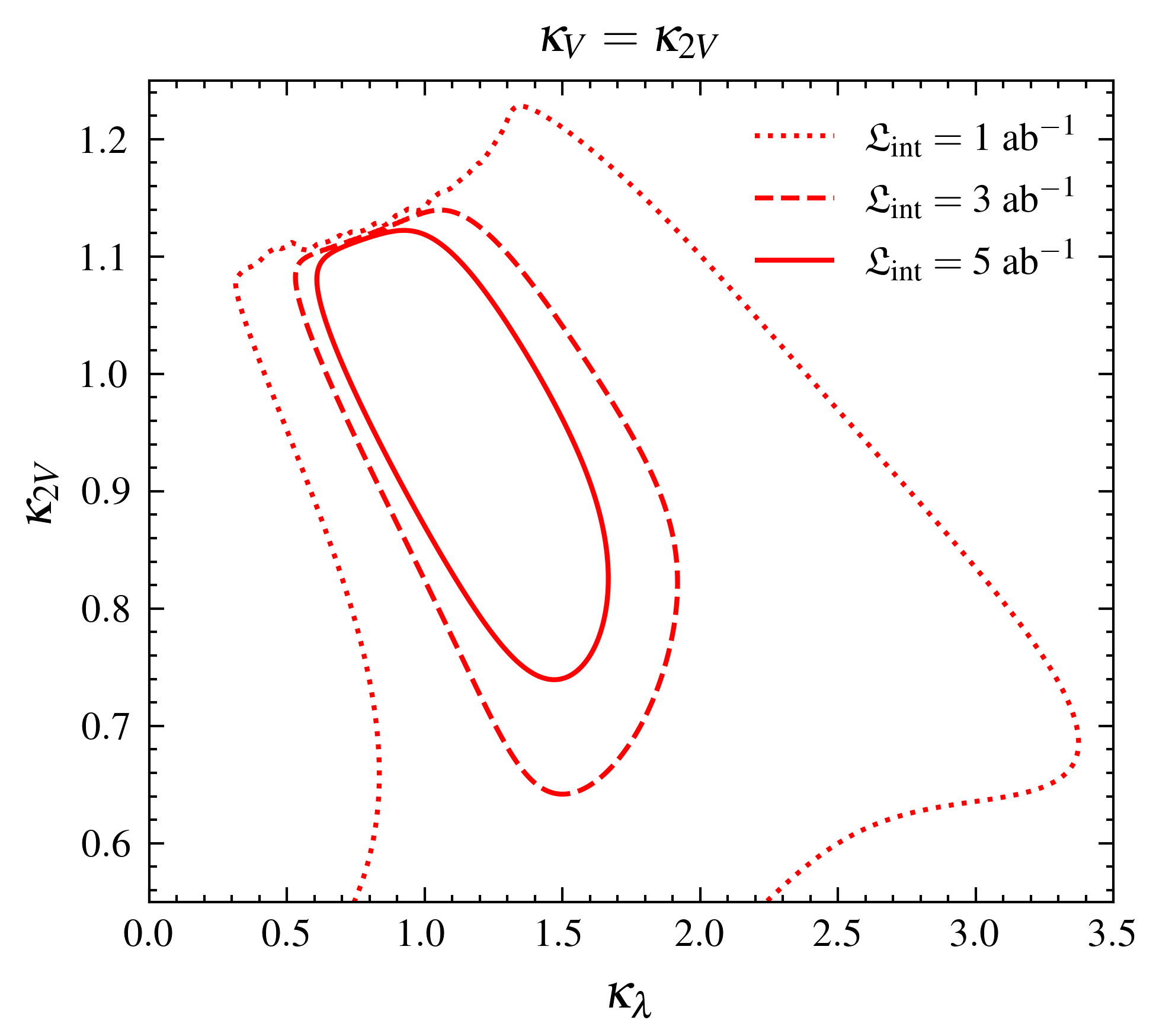}
    \caption{$95\%$ C.L. 2D sensitivity contours at $\mathfrak{L}_{\rm
    int}=1$, 3, and 5~ab$^{-1}$. \textit{Left}: \textit{Case 1}
    ($\kappa_V=1$). \textit{Right}: \textit{Case 2}
    ($\kappa_V=\kappa_{2V}$).}
    \label{fig:lum}
\end{figure}

The corresponding 2D likelihood contours in the $\kappa_{\lambda}$-$\kappa_{2V}$ plane are shown in Fig.~\ref{fig:lum}. For both scenarios the allowed regions contract progressively with
increasing luminosity. In \textit{Case 1}, the contours exhibit a pronounced correlation between $\kappa_\lambda$ and $\kappa_{2V}$ that becomes increasingly localized around the SM point. In \textit{Case 2}, the contours are broader with a different orientation, reflecting the
modified interference pattern from the condition $\kappa_V=\kappa_{2V}$.

To assess the impact of systematic uncertainties, we introduce a single normalized nuisance parameter $\theta$. The systematically shifted prediction in bin $i$ is
\begin{equation}
\mu_i(\kappa,\theta) = \mu_i^0(\kappa)\cdot(1+\sigma_{\rm sys}\,\theta),
\end{equation}
where $\mu_i^0(\kappa)$ is the nominal signal-plus-background prediction and $\sigma_{\rm sys}$ parametrizes the overall systematic impact. The full likelihood is
\begin{equation}
\mathscr{L}(\kappa,\theta)
= \prod_{i}\frac{\mu_i(\kappa,\theta)^{n_i}
  e^{-\mu_i(\kappa,\theta)}}{n_i!}
  \cdot\frac{1}{\sqrt{2\pi}}e^{-\theta^2/2},
\end{equation}
and the physics parameters are extracted via the profile likelihood ratio
\begin{equation}
-2\Delta\ln\mathscr{L}(\kappa)
= \min_\theta\!\left[-2\ln\mathscr{L}(\kappa,\theta)\right]
- \min_{\kappa,\theta}\!\left[-2\ln\mathscr{L}(\kappa,\theta)\right].
\end{equation}

We repeat the analysis for $\sigma_{\rm sys}=0$, $0.05$, and $0.1$ at $\mathfrak{L}_{\rm int}=5~\mathrm{ab}^{-1}$; the resulting 1D limits are given in Table~\ref{tab:sys1}. The sensitivity degrades gradually with increasing systematics, most noticeably for $\kappa_\lambda$, while the limits on $\kappa_{2V}$ remain comparatively robust in both scenarios.

\begin{table*}[htb!]
    \centering
    \caption{1D sensitivity limits at $\mathfrak{L}_{\rm int}=5~\mathrm{ab}^{-1}$
    for systematic uncertainties $\sigma_{\rm sys}=0$, $0.05$, $0.1$.}
    \label{tab:sys1}
    \renewcommand{\arraystretch}{1.3}
    \begin{tabular*}{0.75\textwidth}{@{\extracolsep{\fill}} c c c c c @{}}
    \hline
    $\sigma_{\rm sys}$ & C.L. & $\kappa_{\lambda}$ &
    $\kappa_{2V}$ (\textit{Case 1}) & $\kappa_{2V}$ (\textit{Case 2}) \\
    \hline
    \multirow{2}*{$0$} & $68\%$ & $[0.88,1.14]$ & $[0.98,1.02]$ & $[0.95,1.05]$ \\
     & $95\%$ & $[0.77,1.31]$ & $[0.95,1.05]$ & $[0.90,1.10]$ \\ \hline
    \multirow{2}*{$0.05$} & $68\%$ & $[0.86,1.16]$ & $[0.97,1.03]$ & $[0.94,1.05]$ \\
     & $95\%$ & $[0.73,1.36]$ & $[0.95,1.06]$ & $[0.88,1.10]$ \\ \hline
    \multirow{2}*{$0.1$} & $68\%$ & $[0.82,1.21]$ & $[0.96,1.04]$ & $[0.93,1.06]$ \\
     & $95\%$ & $[0.66,1.47]$ & $[0.93,1.07]$ & $[0.86,1.11]$ \\
    \hline
    \end{tabular*}
\end{table*}

\begin{figure}[htb!]
    \centering
    \includegraphics[width=0.75\columnwidth]{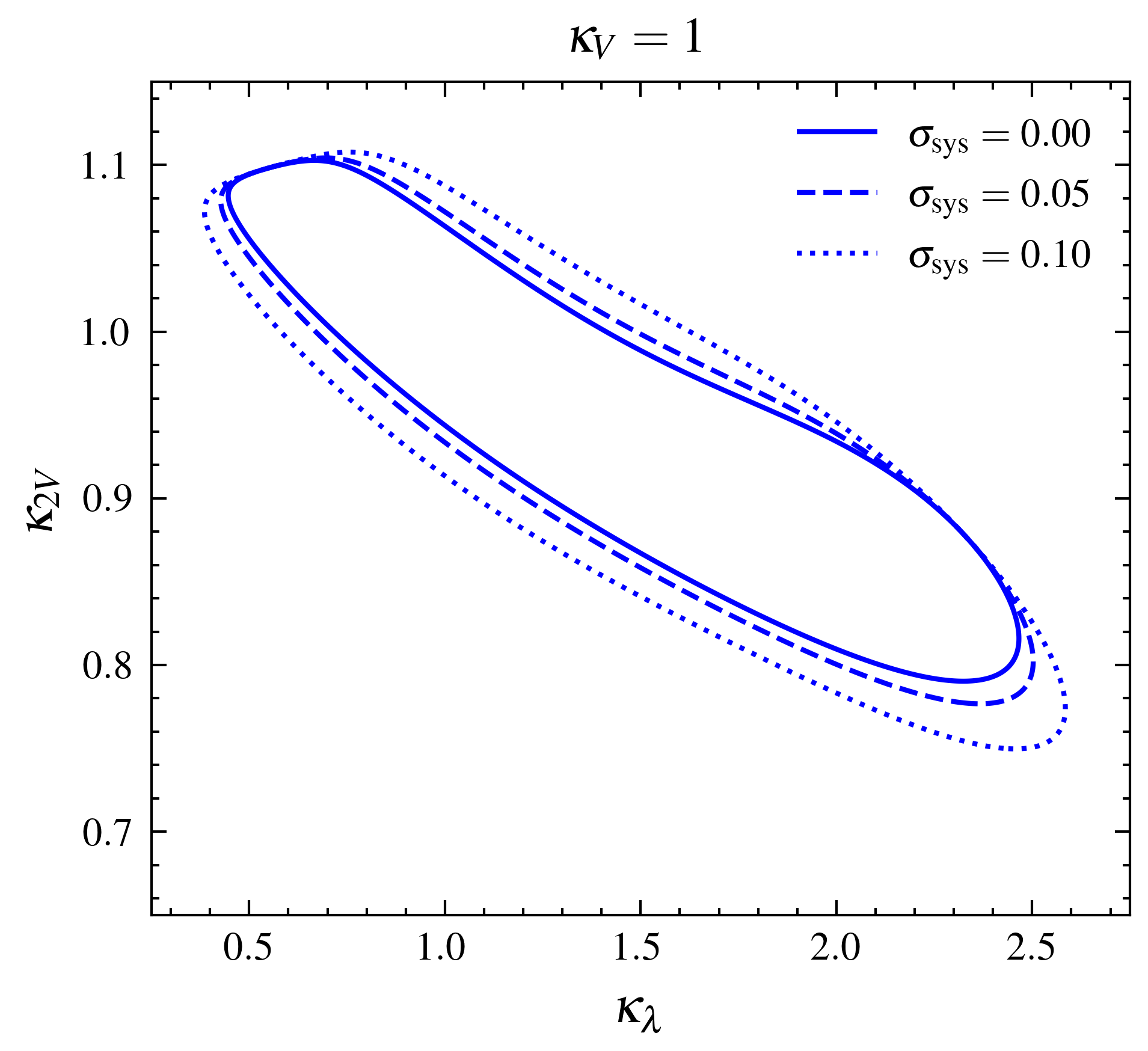}
    \includegraphics[width=0.75\columnwidth]{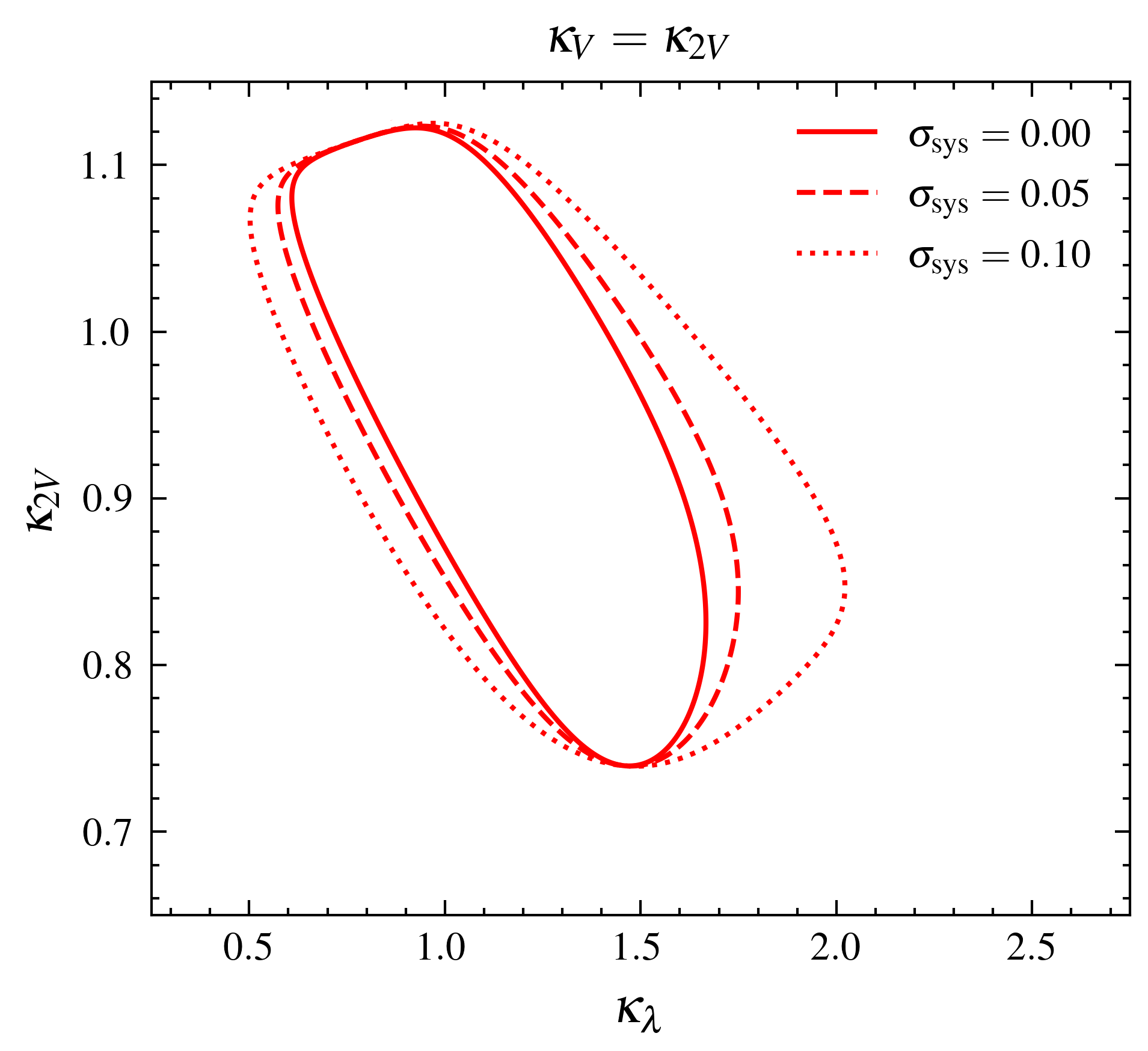}
    \caption{$95\%$ C.L. 2D sensitivity contours at
    $\mathfrak{L}_{\rm int}=5~\mathrm{ab}^{-1}$ for $\sigma_{\rm sys}=0$,
    $0.05$, $0.1$. \textit{Left}: \textit{Case 1} ($\kappa_V=1$).
    \textit{Right}: \textit{Case 2} ($\kappa_V=\kappa_{2V}$).}
    \label{fig:sys}
\end{figure}

The corresponding $95\%$ C.L. 2D contours are shown in Fig.~\ref{fig:sys}. For both scenarios the allowed regions broaden with increasing $\sigma_{\rm sys}$ while preserving their overall shapes, demonstrating that the projected reach of the 3~TeV CLIC remains robust against moderate systematic uncertainties.

\section{Comparison with Boosted Decision Trees}
\label{app:bdt}

To benchmark the performance of the GNN classifier against a well-established multivariate technique, we train a gradient-boosted decision tree on a set of high-level, physically motivated input features built from the reconstructed objects. BDTs remain the standard workhorse of collider analyses owing to their speed, interpretability, and ability to model complex non-linear correlations in tabular data, and therefore provide a natural, strong baseline for the signal-versus-background classification task considered here.

The BDT is trained on a set of features constructed entirely at the detector-simulation level, using only the reconstructed jets and the missing transverse momentum in each event. These features are designed to capture the dominant kinematic differences between the signal $hh\slashed{E}$ and the two reducible backgrounds $hb\bar{b}\slashed{E}$ and $t\bar{t}X$, without relying on any generator-level information. For each event we use:
\begin{enumerate}
    \item $\{p_{T\,j_i},\,\eta_{j_i},\,\phi_{j_i},\,E_{j_i}\}$ with $i=1,\dots,4:$ The four-momentum components of the four hardest jets;
    \item $\Delta\eta_{ij}$, $\Delta\phi_{ij}:$ The pseudorapidity and azimuthal separations between every pair $(i,j)$ of the four leading jets;
    \item $\Delta R_{ij} = \sqrt{(\Delta\eta_{ij})^{2}+(\Delta\phi_{ij})^{2}}:$ The geometric distance between two jets;
    \item $m_{ij} = \sqrt{(p_{i}+p_{j})^{2}}:$ The invariant mass of a pair of jets;
    \item $m_{4j} = \sqrt{(p_{1}+p_{2}+p_{3}+p_{4})^{2}}:$ The invariant mass of the four leading jets;
    \item $p_{T}^{ij} = |\vec{p}_{T\,i} + \vec{p}_{T\,j}|:$ The transverse momentum of a pair of jets;
    \item $\slashed{p}_{T}$ and $\slashed{\phi}:$ The magnitude and azimuthal orientation of the missing transverse momentum vector.
\end{enumerate}

All pairwise quantities are computed for the six unique jet pairs, giving a total of 50 input features per event. Prior to training, every feature is rescaled to the unit interval $[0,1]$, in order to ensure that the tree splits are not biased by the different dynamic ranges of the input variables.

We employ the \textsc{XGBoost}~\cite{Chen_2016} implementation to perform a three-class classification of the signal $hh\slashed{E}$ and the two reducible backgrounds: $hb\bar{b}\slashed{E}$ and $t\bar{t}X$. The raw output of the ensemble for an event $i$ and class $k$ is passed through a softmax function to yield a normalized probability,
\begin{equation}
    p_k(x_i) = \frac{\exp\!\left(\hat{y}_{i,k}\right)} {\sum_{c=1}^{C} \exp\!\left(\hat{y}_{i,c}\right)},
    \qquad C = 3,
\end{equation}
and the model is trained by minimizing the multi-class cross-entropy loss
\begin{equation}
    l\!\left(y_i,\,\hat{y}_i\right) = -\sum_{k=1}^{C} y_{i,k}\,\log p_k(x_i),
    \label{eq:bdt:loss}
\end{equation}
where $y_{i,k}$ is the one-hot encoded true label. Overfitting is controlled through $L_1$ and $L_2$ regularization of the leaf weights, a minimum split-gain threshold, and constraints on tree depth and feature sub-sampling. We use a learning rate of $\eta = 0.01$, $2000$ boosting rounds, a maximum
tree depth of $5$, $L_1$ and $L_2$ penalty coefficients $\alpha = \lambda = 1.5$, a minimum split gain $\gamma = 1.5$, and a feature sub-sampling fraction of $0.9$.

\begin{figure*}[htb!]
    \centering
    \includegraphics[width=0.3\textwidth]{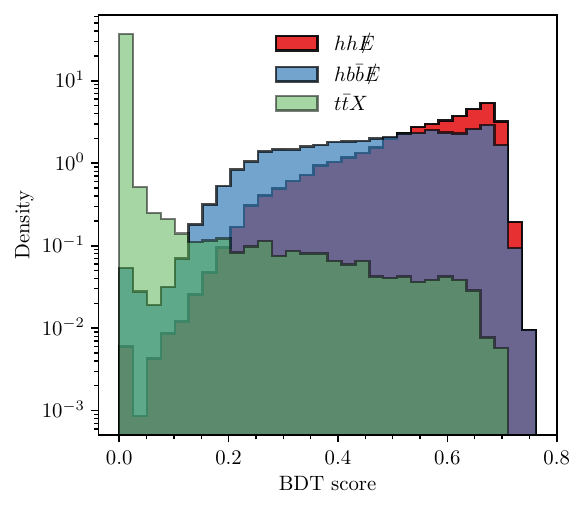}
    \includegraphics[width=0.335\textwidth]{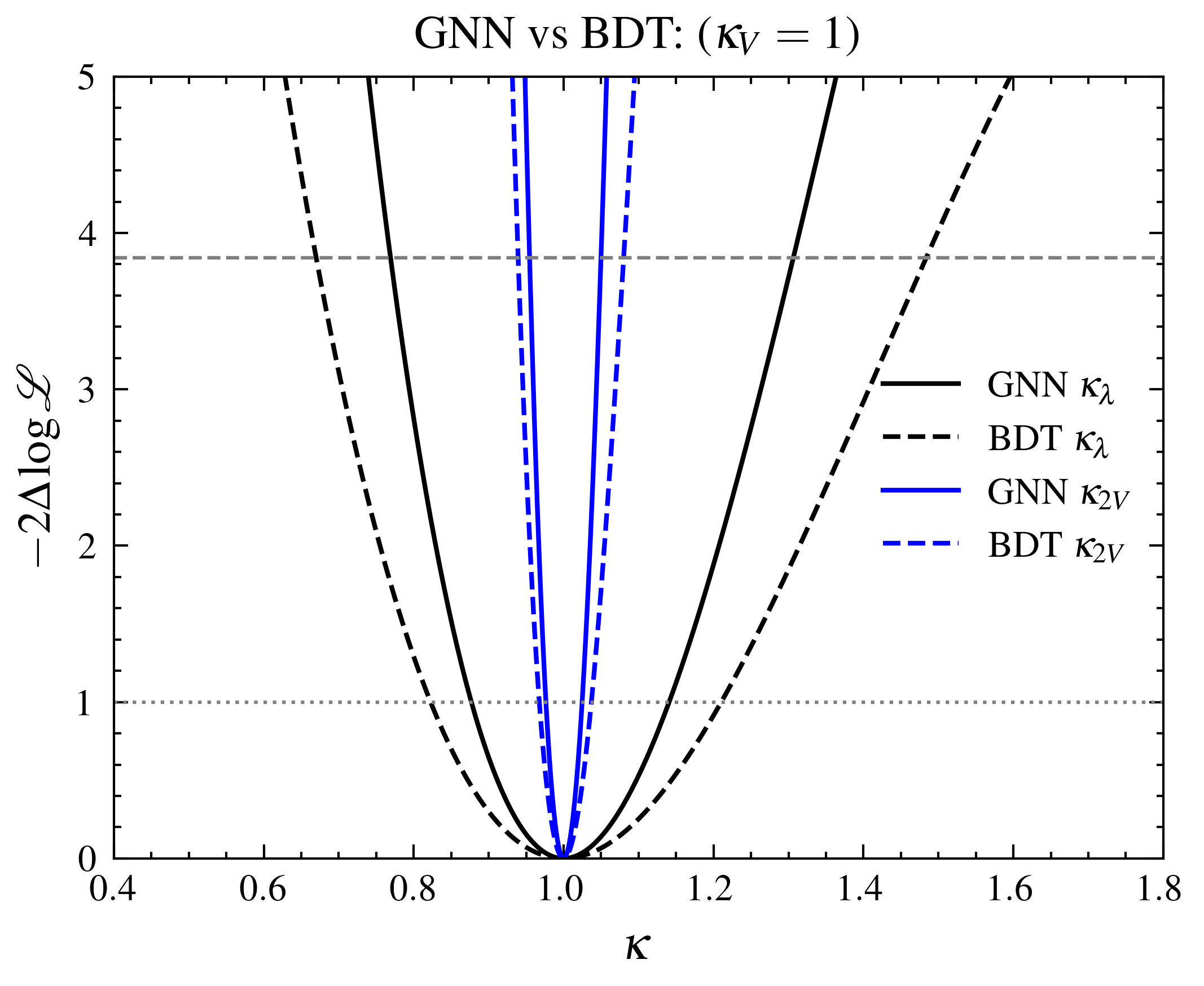}
    \includegraphics[width=0.34\textwidth]{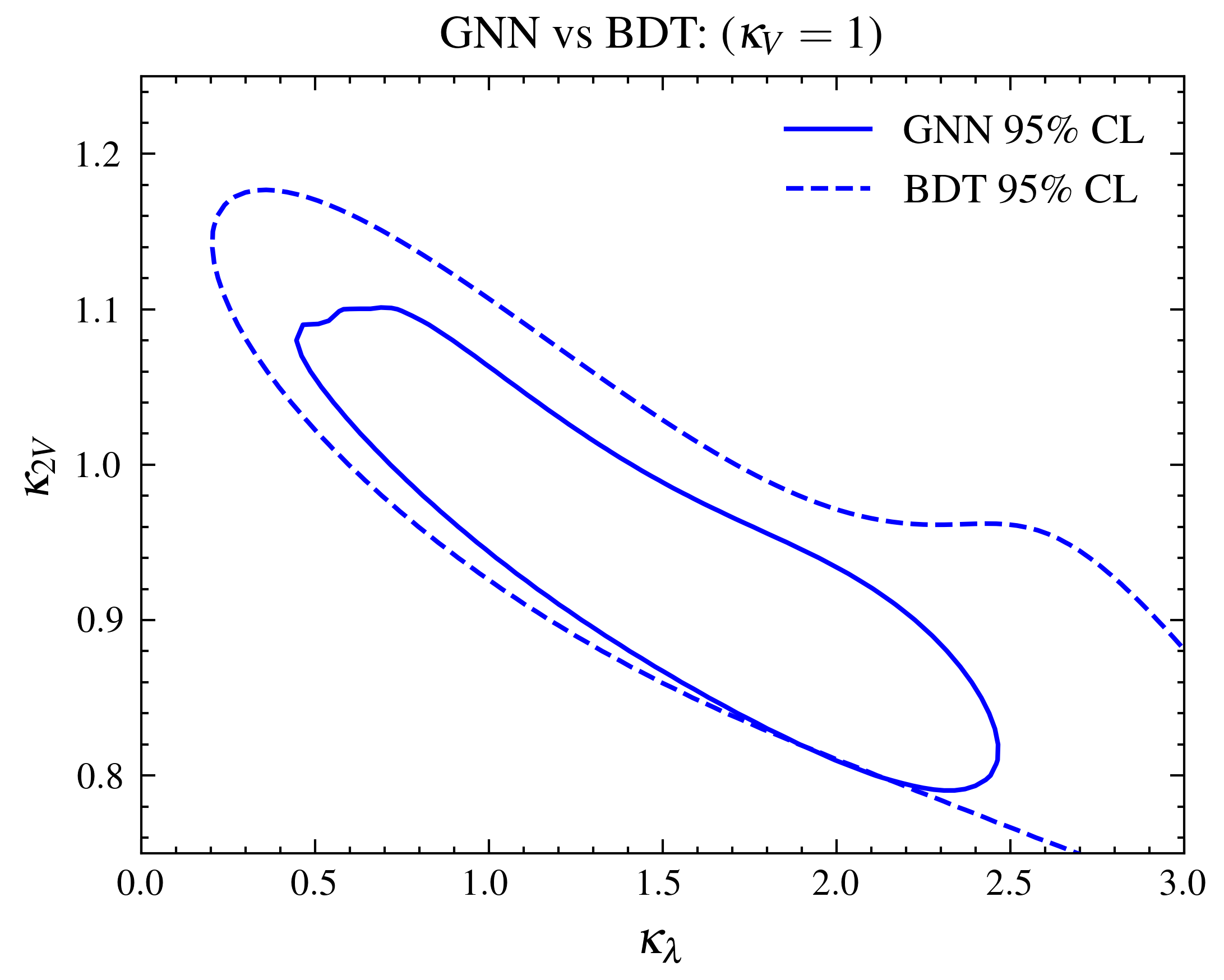}
    \caption{Boosted decision tree output score distribution for the signal $hh\slashed{E}$ and the two backgrounds $hb\bar{b}\slashed{E}$ and $t\bar{t}X$ (\textit{left}). 1D likelihood plots (\textit{center}), and 2D 95\% C.L. sensitivity countours (\textit{right}) for $\kappa_V=1$ case, showing comparison between sensitivity obtained from GNN and BDT classifiers, respectively.}
    \label{fig:bdt_score}
\end{figure*}

The receiver operating characteristic (ROC) curve of the trained BDT yields an area under the curve of $\mathrm{AUC} = 0.77$ for the signal against the two reducible backgrounds, confirming that the high-level kinematic features alone provide a meaningful but not exhaustive discrimination. The
corresponding discriminant output is shown in Fig.~\ref{fig:bdt_score} (\textit{left} panel). The three classes are not sharply separated: while the $t\bar{t}X$ background is clearly peaked towards the low-score region, the $hb\bar{b}\slashed{E}$ background follows a distribution similar to that of the signal, with a
visible distinction emerging only in the high-score tail. We therefore choose a classifier threshold of $0.5$ on the BDT output, which maximize the expected signal significance without a significant loss of signal purity.

We utilize the same likelihood analysis framework that was used for the GNN case in Sec.~\ref{sec:sens} to obtain the sensitivity reach for the BDT case. The 68\% and 95\% C.L. limits on $\kappa_{\lambda}$ are as follows:

\begin{equation}
\kappa_{\lambda}=
\begin{cases}
\text{GNN } \begin{cases}
[0.88,\,1.14] &(68\%~\mathrm{C.L.})\\
[0.76,\,1.31] &(95\%~\mathrm{C.L.})
\end{cases}, \\
\text{BDT } \begin{cases}
[0.82,\,1.21] &(68\%~\mathrm{C.L.})\\
[0.67,\,1.48] &(95\%~\mathrm{C.L.})
\end{cases}.
\end{cases}
\end{equation}

Similarly, for $\kappa_{2V}$, the limits for \textit{Case 1} $(\kappa_V =1)$ are:
\begin{equation}
\kappa_{2V}=
\begin{cases}
\text{GNN } \begin{cases}
[0.98,\,1.02] &(68\%~\mathrm{C.L.})\\
[0.95,\,1.05] &(95\%~\mathrm{C.L.})
\end{cases}, \\
\text{BDT } \begin{cases}
[0.97,\,1.04] &(68\%~\mathrm{C.L.})\\
[0.94,\,1.08] &(95\%~\mathrm{C.L.})
\end{cases}.
\end{cases}
\end{equation}
The limits for \textit{Case 2} $(\kappa_V = \kappa_{2V})$ exhibit a similar trend and, hence, are not explicitly quoted. In Fig.~\ref{fig:bdt_score} (\textit{center} and \textit{right} panels), we present the 1D likelihood plots and 2D 95\% C.L. sensitivity contours, respectively, for the GNN and BDT classifier models. The GNN model significantly improves the sensitivity owing to its ability to effectively eliminate the $t\overline{t}X$ background while retaining a large fraction of the signal events.

\bibliography{refer}
\end{document}